\documentclass[12pt]{article}
\usepackage{amssymb,epsfig}

\begin{document}

\begin{titlepage}

\begin{flushright}
  KUNS-1794\\
  HUPD-0203\\
\end{flushright}

\begin{center}
\LARGE{\bf Schwinger-Dyson Analysis of 
Dynamical Symmetry Breaking on a Brane with 
Bulk Yang-Mills Theory}
\end{center}

\large
\begin{center}
Hiroyuki~Abe$^{*,\dagger,}$\footnote{E-mail address:
    abe@gauge.scphys.kyoto-u.ac.jp}~and~
Tomohiro~Inagaki$^{\S,}$\footnote{E-mail address:
    inagaki@hiroshima-u.ac.jp}
\end{center}

\normalsize
\begin{center}
$^*${\it Department of Physics, Kyoto University, 
Kyoto 606-8502, Japan}\\

$^\dagger${\it Department of Physics, Hiroshima University, 
Hiroshima 739-8526, Japan}\\

$^\S${\it Information Media Center, Hiroshima University, 
Hiroshima 739-8521, Japan}
\end{center}

\begin{abstract}
The dynamically generated fermion mass is investigated in the 
flat brane world with $(4+\delta)$-dimensional bulk space-time, 
and in the Randall-Sundrum (RS) brane world. We consider the 
bulk Yang-Mills theory interacting with the fermion confined 
on a four-dimensional brane. Based on the effective theory below 
the {\it reduced} cutoff scale on the brane, we formulate the 
Schwinger-Dyson equation of the brane fermion propagator. 
By using the improved ladder approximation we numerically
solve the Schwinger-Dyson equation and find that the dynamical 
fermion mass is near the {\it reduced} cutoff scale on the 
brane for the flat brane world with $\delta \ge 3$ and for the 
RS brane world. In RS brane world KK excited modes of the bulk
gauge field localized around the $y=\pi R$ brane and it enhances 
the dynamical symmetry breaking on the brane. The decay 
constant of the fermion and the anti-fermion composite 
operator can be taken to be the order of the electroweak scale 
much smaller than the Planck scale. Therefore electroweak mass
scale can be realized from only the Planck scale in the RS 
brane world due to the fermion and the anti-fermion pair 
condensation. That is a dynamical realization of 
Randall-Sundrum model which solves the weak-Planck hierarchy problem.
\end{abstract}

\end{titlepage}

\section{Introduction}
Up to the energy scale we can test experimentally, the standard model 
(SM) well describes the forces between matter particles except for 
the gravity. Almost all the candidates 
for the theory of the gravity at Planck scale $M_{\rm Pl}$ are defined 
with the supersymmetry (SUSY) in more than 4-dimensional (4D) 
space-time, such as supergravity or superstring theory. 
It is considered that the SM particles lie in higher dimensional 
space-time and extra dimensions are compactified smaller than the 
size we can detect in the low energy experiments or lie on the 4D 
subspace in higher dimensional space-time. 
Recent interesting suggestion is that SM may be realized on a 
`brane' which stands for the lower dimensional object in the 
higher dimensional space-time. 
One of the candidates of such object is D-brane or domain wall. 
We call such scenario `brane world model.' 
The brane world models now casting new ideas on physics beyond the SM. 
For example recent studies in the superstring theory provide us 
some concrete examples realizing SM in the D-brane 
systems \cite{Ibanez:2001nd}. 

Basis of the SM is a spontaneous gauge symmetry breaking 
that results in the existence of the massive gauge bosons, W and Z. 
We need at least one (doublet) scalar field, called Higgs field 
which develops a vacuum expectation value (VEV) to give a masses 
to the W and Z gauge bosons and breaks the SM gauge symmetry. 
These gauge boson masses are related to the electroweak scale 
$M_{\rm EW}$ at which the SM gauge symmetry is broken as 
$SU(3)_c \times SU(2)_L \times U(1)_Y \to SU(3)_c \times U(1)_{EM}$. 
The mass scale of the W and Z gauge bosons shows that $M_{\rm EW}$ is 
of the order TeV scale, while the gravitational scale, namely Planck 
scale $M_{\rm Pl}$ is of the order $10^{19}$ GeV. 
If we consider the unification of SM and gravity, we should 
explain the hierarchical structure between $M_{\rm Pl}$ and $M_{\rm EW}$. 
Since the mass of the scalar field is not protected by any symmetries,
the radiative correction brings the Higgs mass to the fundamental 
scale $M_{\rm Pl}$ without a fine tuning. This is so called `hierarchy' or 
`fine tuning' problem. There are two remarkable brane world models 
solving this hierarchy problem. One is the large extra dimension 
scenario \cite{Antoniadis:1990ew} and the other is the warped extra dimension 
scenario in Randall-Sundrum (RS) brane world \cite{Randall:1999ee}.

SUSY enables to stabilize the scalar mass against the radiative 
correction of ${\cal O}(M_{\rm Pl})$ because it is related to 
the mass of the fermionic superpartner that can be protected 
by appropriate symmetry. So the SUSY permits the 
light scalar field like Higgs field, however if we introduce 
it around the TeV scale in order to keep the Higgs mass $\sim$ TeV, 
all the fermions in the SM have their scalar partners 
with their masses around TeV. These extra scalar fields 
have not been observed in any experiments yet. Thus we need some 
mechanisms in the SUSY breaking process 
to control these extra light scalars not to appear inside 
the region of the present experimental observation. 
However, the supertrace theorem predicts lighter scalar partners 
than the fermions in SM that is not consistent with the experimental 
results. To avoid this problem the theory should have the `hidden sector' 
where SUSY is broken. A certain mediation mechanism to communicate 
it to the `visible sector' is required at the loop level through 
renormalizable interactions or at the tree level through unrenormalizable 
interactions. Recently some people consider that the hidden sector setup 
merges into the brane world picture, that separates the hidden and 
visible sector by spatiality in the extra dimension. They use the bulk 
(moduli) fields to communicate the SUSY breaking signals from 
the hidden to the visible sector \cite{Brignole:1994dj}. 
The hidden sector is the minimal requirement from the supertrace 
theorem and we need more severe conditions to the 
SUSY breaking from the experiment. One of them is from 
the supersymmetric flavor problem, that is the masses of the SM 
fermions are disjointed each other while the masses of their scalar 
partners are almost degenerate. 

As we see above almost all the problems in the TeV scale SUSY come from 
too much light extra fields in addition to the SM one. We need complicated 
setup about SUSY breaking (and its mediation) mechanism to avoid these 
problems. 
Remember that SUSY itself is needed for the consistency of the quantum 
theory of gravity, while `TeV scale SUSY' is required in order to stabilize 
the electroweak scale $M_{\rm EW}$, and to realize gauge coupling 
unification. However, in a recent brane world picture, 
there is a possibility that the strong and electroweak gauge group 
come from the different brane. In this case the unification 
of SM and gravity is able to occur directly without grand unification
and the gauge coupling unification in MSSM may be accidental. 
We can also consider the case that gauge coupling unification 
is not depend on the TeV scale SUSY (MSSM) and it happens by the 
other mechanism, e.g. extra dimensional effect 
\cite{Dienes:1998vh,Randall:2001gc}. 
If there is another mechanism to stabilize mass of the Higgs field, 
it is possible to break SUSY at higher energy scale, even around 
the Planck scale. Such a mechanism can set SUSY free from any 
problems at TeV scale.

In this paper we notice the recent suggestions that the electroweak 
(chiral) symmetry is broken down dynamically by the gauge interaction
in more than 4D space-time 
\cite{Dobrescu:1999dg,Rius:2001dd,Hashimoto:2001uk,Abe:2000ny,Abe:2001yi}. 
The higher dimensional gauge theory may realize the stabilization of 
a light Higgs field through the composite Higgs scenario. This is based 
on the idea that the gauge theory in compact extra dimension has 
Kaluza-Klein (KK) massive gauge bosons and they act as a binding force 
between fermion and anti-fermion that results in the composite scalar 
field. It was pointed out in Refs.~\cite{Inagaki:1993ya} 
and \cite{Inagaki:1995jp} a negative curvature and a finite size 
effects enhance dynamical symmetry breaking\footnote{SUSY extension of 
these models are also interesting \cite{Buchbinder:1997ta}.}. 
If SM is embedded in a brane world, the dynamical mechanism of 
electroweak symmetry breaking can be realized in a certain bulk gauge 
theory (or SM itself in the bulk.) 
SM in the bulk has a possibility to realize 
top quark condensation scenario \cite{Miransky:1988xi} in the brane world. 
It is very interesting because it may have no extra field (elementary 
Higgs, techni-color gauge boson, techni-fermion etc.) except for the 
SM contents (without elementary Higgs) in order to break the
electroweak symmetry. We launched a plan to study the dynamical 
symmetry breaking in the brane world models in detail.
 
There is no enough knowledge about the (non-perturbative) 
dynamics of the higher dimensional gauge theory in brane world. 
In this paper we study the basic properties of dynamical symmetry 
breaking in the bulk gauge theory couples to a fermion on the brane. 
Based on the effective theory of the bulk Yang-Mills theory
that is defined below the {\it reduced} cutoff scale on the brane, 
we formulate the ladder Schwinger-Dyson (SD) equation of the fermion 
propagator. It is numerically solved by using the improved ladder 
approximation with power low running of the effective gauge coupling.
We show how the dynamics of the bulk gauge field affects
the chiral symmetry breaking on the brane. 

We give a KK reduced Lagrangian of our bulk-brane system in $\S$\ref{sec:lag}. 
In $\S$\ref{sec:effsd} we derive the effective theory on the four-dimensional 
brane and formulate SD equation about the fermion propagator on the brane. 
The Numerical analysis of the SD equation is performed, within the improved 
ladder approximation, for the case of Yang-Mills theory on the brane, 
in the flat bulk space-time and in the warped bulk space-time in $\S$\ref{sec:ila}.

\section{Lagrangian of the System} \label{sec:lag}

In this paper we analyze dynamical symmetry 
breaking on the four dimensional brane with bulk gauge theory 
in various brane world models. The bulk gauge theory is, of course, 
more than four dimensional theory and ill defined in ultraviolet (UV) 
region and we need some regularization about it. 
In this paper we use effective Lagrangian of it defined below the 
{\it reduced} cutoff scale on the brane. 

In this section we derive the KK reduced 4D Lagrangian of the bulk 
gauge theory in the brane worlds, $M_4 \times S^1$ type space-time. 
We can easily extend it to the case of higher dimensional 
bulk space-time, $M_4 \times T^\delta$, it is shown in the next section. 
Because one of our goal is to analyze the dynamically induced mass 
scale on the brane in the Randall-Sundrum space-time, 
i.e. the slice of AdS$_5$, 
we take account for the Lagrangian and mode function in a curved 
extra dimension. The original motivation to introduce a curved extra 
dimension by Randall and Sundrum is to produce the weak-Planck hierarchy 
from the exponential factor in the space-time metric. 
The factor is called `warp factor', and the extra space `warped 
extra dimension'. A non-factorizable geometry with the warp factor 
distinguishes the RS brane world from the others. 
In the RS model we consider the fifth dimension $y$ which is compactified 
on an orbifold, $S^1/Z_2$ of radius $R$ and two 3-branes at the orbifold 
fixed points, $y^\ast=0$ and $\pi R$. 
Requiring the bulk and boundary cosmological constants to be related, 
Einstein's equation in five-dimension leads to the solution 
\cite{Randall:1999ee}, 
\begin{equation}
ds^2=G_{MN}dx^Mdx^N=e^{-2k|y|}\eta_{\mu\nu}dx^\mu dx^\nu - dy^2, 
\label{eq:RSmetric}
\end{equation}
where $M,N$ = $0,1,2,3,4$, $\mu,\nu=0,1,2,3$ and 
$\eta_{\mu\nu}={\rm diag}(1,-1,-1,-1)$. 
$k$ is the AdS curvature scale with the mass dimension one, 
and $k=0$ describes the flat $M_4 \times S^1$ space-time. 
In the following we study a bulk vector field, $A_M$ 
in this background.  

\subsection{KK Reduced Lagrangian of Bulk Gauge Theory}
Substituting the metric $G_{MN}$ defined in Eq.~(\ref{eq:RSmetric}) 
and taking $A_4=0$ gauge, the Lagrangian of a bulk gauge field 
in the RS brane world is written by 
\begin{eqnarray}
{\cal L}_{\rm gauge}^{(5)}
= -\frac{1}{4} \sqrt{-G} F^{MN}F_{MN} 
= -\frac{1}{4} F^{\mu \nu}F_{\mu \nu}
    +\frac{1}{2} e^{-2k|y|} (\partial_4 A_\mu)^2
    +{\cal L}_{\rm SI}^{(5)},
\label{eq:bulkgauge}
\end{eqnarray}
where ${\cal L}_{\rm SI}^{(5)}$ includes the gluon three and four point 
self interaction in the bulk. 
We drop the explicit symbol of the trace operation in terms of the 
Yang-Mills index in the Lagrangian throughout this paper.   
We perform the KK mode expansion as 
\begin{eqnarray}
A_\mu (x,y) 
= \frac{1}{\sqrt{2 \pi R}}
  \left(
  \sum_{n=0}^\infty A_\mu^{(n)}(x) \chi_n (y)
 +  \sum_{n=1}^\infty \tilde{A}_\mu^{(n)}(x) \tilde{\chi}_n (y)
  \right), 
\end{eqnarray}
where $\chi_n (y)$ and $\tilde{\chi}_n (y)$ are the $Z_2$ even and odd 
mode function of $n$-th KK excited mode $A_\mu^{(n)}(x)$ and 
$\tilde{A}_\mu^{(n)}(x)$, respectively. 
The $Z_2$ orbifold condition thus means $\tilde{A}_\mu^{(n)}(x) \equiv 0$ 
for $^\forall n$, i.e. dropping all the odd modes. After KK expansion, 
we integrate the Lagrangian (\ref{eq:bulkgauge}) over the y-direction and obtain 
\begin{eqnarray}
{\cal L}_{\rm gauge}
&=& \sum_{n=0}^\infty \left[ 
    -\frac{1}{4}F^{(n)\mu \nu}(x)F_{\mu \nu}^{(n)}(x)
    +\frac{1}{2} M_n^2 A^{(n)\mu} A_{\mu}^{(n)} \right] \nonumber \\ &&
   +\sum_{n=1}^\infty \left[ 
    -\frac{1}{4}\tilde{F}^{(n)\mu \nu}(x)\tilde{F}_{\mu \nu}^{(n)}(x)
    +\frac{1}{2} M_n^2 \tilde{A}^{(n)\mu} \tilde{A}_{\mu}^{(n)} \right], 
\label{eq:bulkgaugekinmass}
\end{eqnarray}
where the KK mode function $\chi_n(y)$ and mass eigenvalue $M_n$ satisfy 
\begin{eqnarray}
-\partial_y \left( e^{-2k|y|} \partial_y \chi_n(y) \right)
  &=& M_n^2 \chi_n(y), \label{eq:KKeigeneq} \\
\frac{1}{2\pi R} \int_{-\pi R}^{\pi R} dy\, \chi_n(y) \chi_m(y) 
  &=& \delta_{nm}, \label{eq:normcond}
\end{eqnarray}
and $\tilde{\chi}_n(y)$ obeys the same equations. 
The boundary conditions are given by 
\begin{eqnarray}
\partial_y \chi_n(y^\ast) = \tilde{\chi}_n(y^\ast) = 0 
\qquad (y^\ast=0,\pi R). 
\label{eq:boundarycond}
\end{eqnarray}
Eq.~(\ref{eq:bulkgaugekinmass}) is the 4D effective Lagrangian 
of the bulk gauge theory and 
Eqs.~(\ref{eq:KKeigeneq})-(\ref{eq:boundarycond})
determine the form of the mode function and the eigenvalue $M_n$. 
Below we solve Eq.~(\ref{eq:KKeigeneq}) with the boundary 
condition (\ref{eq:boundarycond}) and the normalization 
(\ref{eq:normcond}) for the flat ($k=0$) and warped ($k \ne 0$) extra
dimensions. 

We call the case with $k=0$ `flat brane world'. In this case the solution 
of Eqs.~(\ref{eq:KKeigeneq})-(\ref{eq:boundarycond}) is given by
\begin{eqnarray}
\chi_0 (y) &=& 1, \\
\chi_n (y) &=& \sqrt{2} \cos (n y/R) 
  \quad (n =1,2,\ldots), \label{eq:KKwf} \\
\tilde{\chi}_n (y) &=& \sqrt{2} \sin (n y/R) 
  \quad (n =1,2,\ldots), \label{eq:oKKwf} \\
M_n &=& n\mu_R, \label{eq:KKmasstorus}
\end{eqnarray}
where $\mu_R=1/R$. 
These are the KK mode function and mass eigenvalue of the bulk gauge field 
in the flat brane world with $M_4 \times S^1$. 
We can easily extend it to higher dimensional bulk space-time,
$M_4 \times T^\delta$, that is shown latter. 

$k \ne 0$ with orbifold condition, $\tilde{A}_\mu^{(n)}(x) \equiv 0$ 
for $^\forall n$, gives the RS brane world. 
The solution of Eqs.~(\ref{eq:KKeigeneq})-(\ref{eq:boundarycond}) 
reads \cite{Chang:1999nh}
\begin{eqnarray}
\chi_0(y) &=& 1, \\
\chi_n(y) &=& \frac{e^{k|y|}}{N_n} 
\left[ J_1 \left( \frac{M_n}{k} e^{k|y|} \right) 
 + c_n Y_1 \left( \frac{M_n}{k} e^{k|y|} \right) \right] 
  \quad (n=1,2,\ldots), \label{eq:rsKKwf} 
\end{eqnarray}
where the coefficient and the normalization factor are 
given by 
$c_n = -J_0 \left( \frac{M_n}{k} \right)/Y_0 \left( \frac{M_n}{k} \right)$ 
and $N_n^2 = \frac{1}{2 \pi R} \int_{-\pi R}^{\pi R} dy \, 
e^{2k|y|} \left[ J_1 \left( \frac{M_n}{k} e^{k|y|} \right) 
 + c_n Y_1 \left( \frac{M_n}{k} e^{k|y|} \right) \right]^2$ respectively 
and $J$ and $Y$ are the Bessel functions. 
The KK mass eigenvalue $M_n$ is obtained as the solution of 
$J_0 \left( \frac{M_n}{k} \right) 
Y_0 \left( \frac{M_n}{k} e^{\pi kR} \right)
= Y_0 \left( \frac{M_n}{k} \right) 
J_0 \left( \frac{M_n}{k} e^{\pi kR} \right)$. 
For the case with $kR \gg 1$ and $M_n \ll k$ the asymptotic form 
of the KK mass eigenvalue is simplified to
\begin{eqnarray}
M_0 &=& 0, \nonumber \\
M_n &\simeq& (n-1/4)\mu_{kR} \quad (n=1,2,\ldots), 
\label{eq:KKmassasym}
\end{eqnarray}
where $\mu_{kR} \equiv \pi e^{-\pi k R} k$. 
We find that the spectrum of the excited mode is shifted by the 
factor $1/4$ and the mass difference between neighbour modes 
is suppressed by the warp factor $e^{-\pi k R}$. 
In the same asymptotic limit $n$-dependence of the mode function 
vanishes on the $y=\pi R$ brane. It reduces to 
\begin{eqnarray}
\chi_n(\pi R) \simeq \sqrt{2\pi kR}. 
\end{eqnarray}

\begin{figure}[t]
\begin{center}
\begin{minipage}{0.48\linewidth}
   \centerline{\epsfig{figure=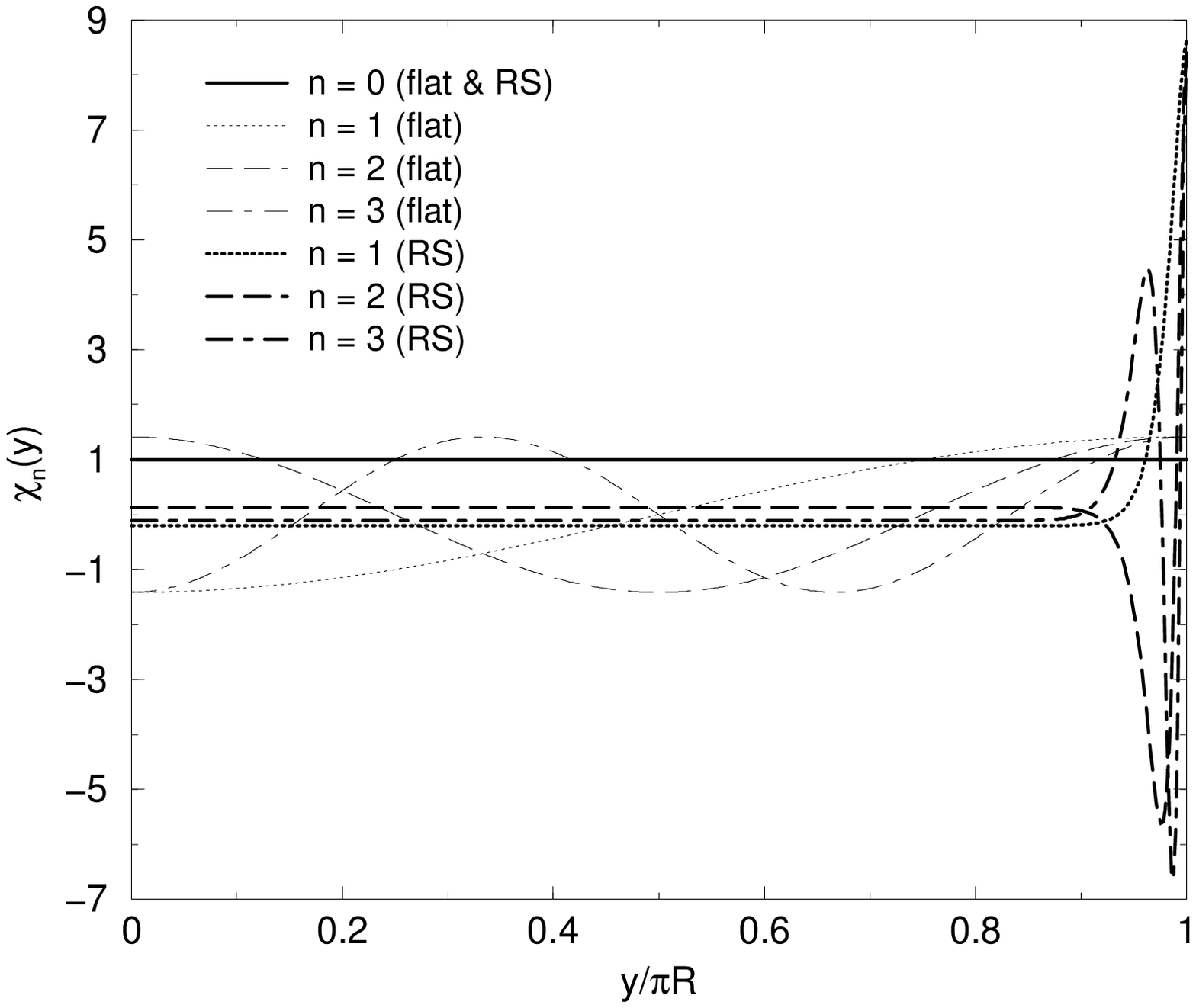,width=\linewidth}}
\end{minipage}
\hfill
\begin{minipage}{0.48\linewidth}
   \centerline{\epsfig{figure=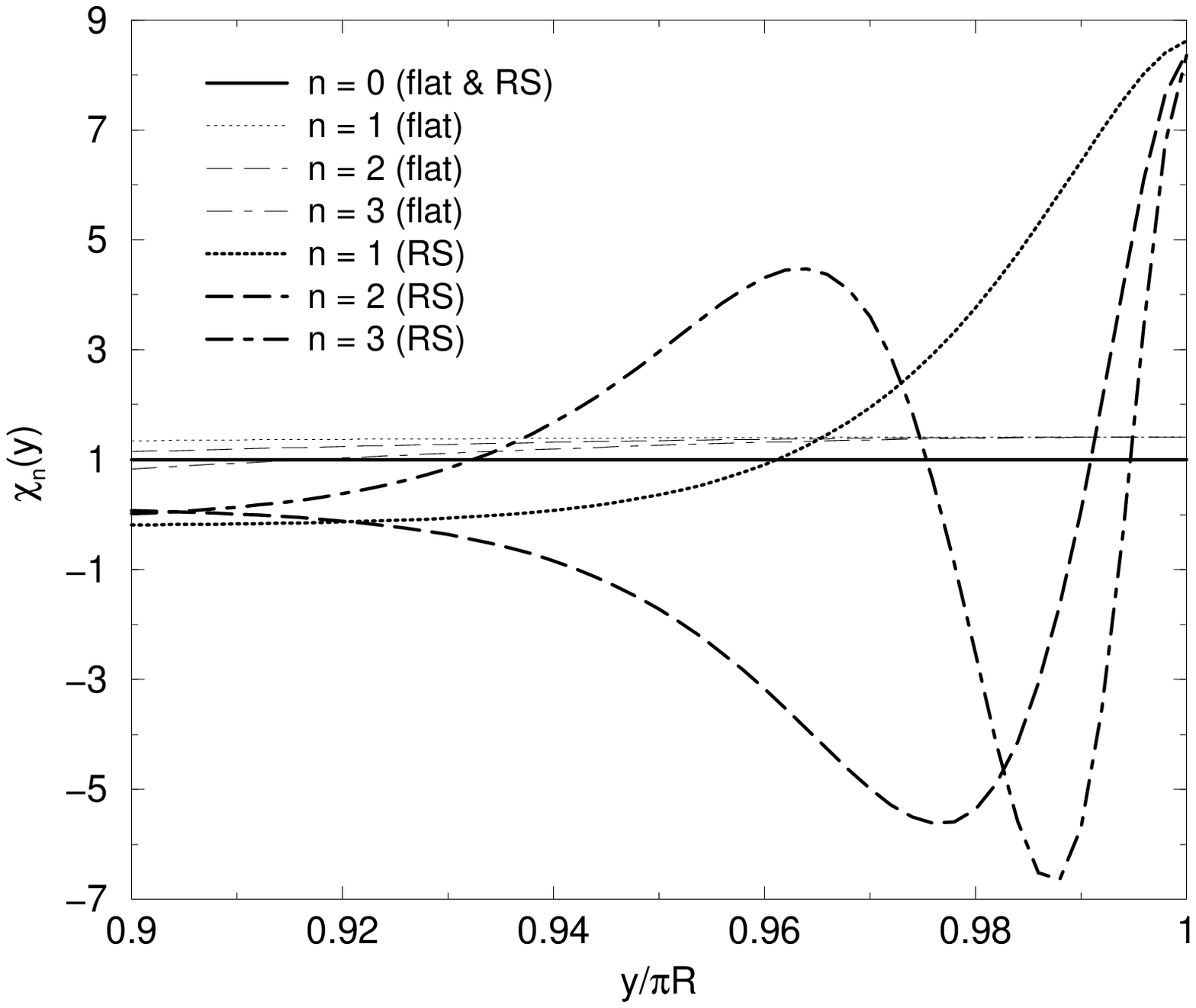,width=\linewidth}}
\end{minipage}
\end{center}
\caption{The profile of the mode functions $\chi_n(y)$ 
for $kR=11$, 
as a function of the coordinate of the extra dimension $y$. 
$\chi_n (y)$ are plotted so as to take a positive value at 
the $y/\pi R =1$. }
\label{fig:WF}
\end{figure}

The profile of the mode functions $\chi_n(y)$ 
are shown in Fig.~\ref{fig:WF}. 
From the figure we know that the 
KK excited gauge bosons are localized in the vicinity of 
the $y=\pi R$ brane in the RS brane world due to the 
finite curvature scale $k$ of the extra dimension, while 
the zero mode is flat in the direction of the extra dimension. 
We will see that the localized KK excited modes 
enhance (suppress) the dynamical symmetry breaking on the $y=\pi R$ 
($y=0$) brane.

\subsection{Coupling to Brane Fermion}

Next we define the coupling of the bulk gauge field 
to a fermion on the brane. By decomposing the metric as 
$G_{MN}=E_{\ M}^{\bar{M}} \eta_{\bar{M}\bar{N}} E_{\ N}^{\bar{N}}$
where 
$\eta_{\bar{M}\bar{N}}={\rm diag}(1,-1,-1,-1,-1)$, 
we define 
\begin{eqnarray}
E_{\bar{M}}^{\ M} = \left(
\begin{array}{cc}
e^{+k|y|}\delta_{\bar{\mu}}^{\ \mu} & 0 \\
0 & 1 
\end{array}
\right), \quad
E_{\ M}^{\bar{M}} = \left(
\begin{array}{cc}
e^{-k|y|}\delta^{\bar{\mu}}_{\ \mu} & 0 \\
0 & 1 
\end{array}
\right), 
\end{eqnarray}
where the row and column of the matrix correspond to the indices 
$M$ and $\bar{M}$ respectively. $E_{\bar{M}}^{M}$ gives the coupling 
of a fermion on the brane and the bulk gauge field. 
That is described as 
\begin{eqnarray}
{\cal L}_{\rm coupling}^{(5)} 
&=& {(\det E_{\ \mu}^{\bar{\mu}})}^{-1} g_{\rm 5D} \bar{\Psi}(x) 
    i\Gamma^{\bar{M}} E_{\bar{M}}^{\ M}(y) A_M(x,y) \Psi(x) 
    \delta (y-y^\ast) \nonumber \\
&=& g_{\rm 5D} \bar{\Psi}(x)
    iA\!\!\!/(x,y) \Psi(x) \delta (y-y^\ast). 
\label{eq:bbcoup}
\end{eqnarray}
where $\Gamma_{\bar{M}}=(\gamma_\mu,i\gamma_5)$, 
$A\!\!\!/ = \gamma^\mu A_\mu$, 
$g_{\rm 5D}$ is the five dimensional gauge coupling and 
$y^\ast$ is the position of the brane. 
In the second line we take $A_4=0$ gauge and 
rescale $\Psi$ field as $\Psi \to e^{3k|y^\ast|/2}\Psi$ 
in order to canonically normalize the kinetic term of $\Psi$. 
By integrating (\ref{eq:bbcoup}) over the extra dimensional 
coordinate we obtain the 4D effective gauge coupling written by 
\begin{eqnarray}
{\cal L}_{\rm coupling} 
&=& \sum_{n=0}^\infty g_n 
    \bar{\Psi}(x) i{A\!\!\!/}^{(n)}(x) \Psi(x) 
   +\sum_{n=1}^\infty \tilde{g}_n 
    \bar{\Psi}(x) i{\tilde{A}\!\!\!/}^{(n)}(x) \Psi(x), 
    \label{eq:4Dgcterm}
\end{eqnarray}
where 
\begin{eqnarray}
g_n = \frac{\chi_n(y^\ast)}{\chi_0(y^\ast)} g_0, \quad 
g_0 = \frac{\chi_0(y^\ast)}{\sqrt{2\pi R}} g_{\rm 5D} 
    \equiv g, 
\label{eq:gn}
\end{eqnarray}
where $g$ is the ordinary four dimensional gauge coupling. 
$\tilde{g}_n$ is defined by replacing $\chi$ with $\tilde{\chi}$ 
in Eq.~(\ref{eq:gn}). 
In the case $y^\ast =0,\pi R$, i.e. $\tilde{g}_n=0$  
(see Eq.~(\ref{eq:boundarycond})), 
$\tilde{A}_\mu^{(n)}(x)$ decouples in Eq.~(\ref{eq:4Dgcterm}). 
We notice that the coupling of the gauge boson 
excited modes $g_n$ are enhanced/suppressed by the value of the 
mode function $\chi_n$ on the $y=y^\ast$ brane, compared with it of 
zero mode. Thus we obtain larger effect from the KK excited gauge 
boson on the brane they localize.

\section{Effective Theory and Schwinger-Dyson Equation on a Brane}
\label{sec:effsd}

In the previous section we have derived the Lagrangian of the 
bulk gauge field and its coupling to a fermion on the brane. 
Putting them together we obtain the effective Lagrangian for the system 
of the bulk gauge theory couples to fermion on the $y=y^\ast(=0,\pi R)$ 
brane as 
\begin{eqnarray}
{\cal L} 
&=& \frac{1}{2} \sum_{n=0}^{N_{\rm KK}} A_\mu^{(n)} \left[ 
    \eta^{\mu \nu} \left( \partial^2 + M_n^2 \right)
   -(1-\xi) \partial_\mu \partial_\nu \right] A_\nu^{(n)}
    \nonumber \\ &&
   +\frac{1}{2} \sum_{n=1}^{N_{\rm KK}} \tilde{A}_\mu^{(n)} \left[ 
    \eta^{\mu \nu} \left( \partial^2 + M_n^2 \right)
   -(1-\xi) \partial_\mu \partial_\nu \right] \tilde{A}_\nu^{(n)}
   +{\cal L}_{\rm SI} \nonumber \\ &&
   +\bar{\Psi} \left( i\partial_\mu + g A_\mu^{(0)} \right) 
    \gamma^\mu \Psi
   + g \sum_{n =1}^{N_{\rm KK}} \frac{\chi_n(y^\ast)}{\chi_0(y^\ast)} 
    \bar{\Psi} A_\mu^{(n)} \gamma^\mu \Psi, 
\label{eq:efflag}
\end{eqnarray}
where ${\cal L}_{\rm SI}$ represents the self interaction (and mixing) 
terms of $A_\mu^{(n)}$ and $\tilde{A}_\mu^{(n)}$, and 
$N_{\rm KK}$ stands for the cutoff of the KK summation. 
It should be noted that we introduce the gauge fixing parameter 
$\xi$ on the brane. 
(See appendix in Ref.~\cite{Abe:2001yi} for more details.) 
$\xi$ is chosen so that a QED like Ward-Takahashi identity 
is hold in the latter analysis. 
We can easily extend the above system to 
higher dimensional bulk space-time with $M_4 \times T^\delta$ 
by replacing $M_n$ and $\chi_n$ as 
\begin{eqnarray}
M_n&\to&\sqrt{M_{n_1}^2+M_{n_2}^2+\ldots+M_{n_\delta}^2},\\
\chi_n(y)&\to&\chi_{n_1}(y_1)\chi_{n_2}(y_2) \cdots \chi_{n_\delta}(y_\delta),
\end{eqnarray}
where $n_1, n_2, \ldots, n_\delta = 0,1,2,\ldots$ 
correspond to the indices of the KK excited mode for each extra dimension 
$(y_1, y_2, \ldots, y_\delta)$ respectively. 
$M_{n_i}$ and $\chi_{n_i}(y_i)$ $(i=1,...,\delta)$ are given 
in Eqs.~(\ref{eq:KKwf}) and (\ref{eq:KKmasstorus}) respectively. 
In addition we should interpret $\sum_n$ as $\sum_{n_1} \sum_{n_2} 
\cdots \sum_{n_3}$ with a cut-off condition, 
$n^2 = n_1^2+n_2^2+\ldots+n_\delta^2 \le N_{\rm KK}^2$. 

In the following discussion we regard the KK reduced Lagrangian 
(\ref{eq:efflag}) as an effective one below the bulk fundamental scale 
$\Lambda_{\rm 5D}$. Thus we cutoff the momentum of the loop integral,
including only the propagator of bulk fields, at the scale 
$\Lambda_{\rm 5D}$. On the other hand we cutoff the momentum
at the {\it reduced} cutoff scale $\Lambda$ on the brane in 
the case that the loop integral includes the brane fields.
Loop integrals in the ladder Schwinger-Dyson equation is only
the latter case. In the flat brane world $\Lambda$ should be 
the same order of $\Lambda_{\rm 5D}$, while in the RS brane world 
we have an large difference between $\Lambda$ and $\Lambda_{\rm 5D}$
due to the warp factor. 
We also define $N_{\rm KK}$ as the number of the KK modes with their mass 
below the {\it reduced} cutoff scale $\Lambda$. 
In our numerical analyses, however, we cutoff the KK mode summation 
at $N_{\rm KK}+10$ in order to avoid the sharp threshold effect.

From the Lagrangian (\ref{eq:efflag}) we construct a Schwinger-Dyson 
equation for the fermion propagator, 
\begin{eqnarray}
iS^{-1}(p) 
= iS^{-1}_0(p) + \sum_{n=0}^{N_{\rm KK}} \int \frac{d^4 q}{i(2\pi)^4}
  \left[ -ig_n T^a \Gamma^M \right] S(q) 
  \left[ -ig_n T^a \Gamma^N \right] D_{MN}^{(n)} (p-q), 
\label{eq:orgSDeq}
\end{eqnarray}
where $T^a$ is the gauge group representation matrix of the fermion, and 
$S_0(p) = i/p\!\!\!/$, $S(p)$ and $D_{MN}^{(n)} (p-q)$ are the free fermion 
propagator, full fermion propagator and full $n$-th excited KK gauge boson 
propagator respectively. In this paper we use the (improved) ladder 
approximation in which $D_{MN}^{(n)}$ is replaced by the free propagator, 
\begin{eqnarray}
D_{MN}^{(n)}(k) &\equiv& \frac{-i}{k^2-M_n^2}
\left[ \eta_{MN} -(1-\xi)\frac{k_Mk_N}{k^2-\xi M_n^2} \right]. 
\end{eqnarray}

By writing the full fermion propagator as 
\begin{eqnarray}
iS^{-1} (p) &\equiv& A(-p^2) p\!\!\!/ -B(-p^2), 
\end{eqnarray}
the SD equation (\ref{eq:orgSDeq}) becomes the 
simultaneous integral equation of $A$ and $B$,
\begin{eqnarray}
A(p^2) &=& 1+ \int_0^{\Lambda^2} dq^2\, 
            \frac{q^2A(q^2)}{A^2(q^2)q^2 +B^2(q^2)} 
            \sum_{n=0}^{N_{\rm KK}} L_\xi (p^2,q^2;M_n,\alpha_n),  
\label{eq:SDA} \\
B(p^2) &=& \int_0^{\Lambda^2} dq^2\, 
         \frac{q^2 B(q^2)}{A^2(q^2)q^2 +B^2(q^2)} 
         \sum_{n=0}^{N_{\rm KK}} K_\xi (p^2,q^2;M_n,\alpha_n), 
\label{eq:SDB}
\end{eqnarray}
where 
\begin{eqnarray}
L_\xi (p^2,q^2;M_n,\alpha_n)
&=& \frac{\alpha_n}{4\pi} 
    \Bigg[ q^2f_{M_n^2}^2 (p^2,q^2) 
         + \frac{2q^2}{M_n^2} \left\{ f_{M_n^2} (p^2,q^2)-
                             f_{\xi M_n^2} (p^2,q^2) \right\} 
           \nonumber \\  && \hspace{1cm} 
         - \frac{(q^2)^2+p^2q^2}{2M_n^2} \left\{ f_{M_n^2}^2 (p^2,q^2)-
                             f_{\xi M_n^2}^2 (p^2,q^2) \right\} 
    \Bigg], \label{eq:kerL} \\
K_\xi (p^2,q^2;M_n,\alpha_n)
&=& \frac{\alpha_n}{4\pi} 
    \bigg[ 4f_{M_n^2} (p^2,q^2) 
         + \frac{p^2+q^2}{M_n^2} \left\{ f_{M_n^2} (p^2,q^2)-
                             f_{\xi M_n^2} (p^2,q^2) \right\} 
           \nonumber \\  && \hspace{1cm} 
         - \frac{p^2q^2}{M_n^2} \left\{ f_{M_n^2}^2 (p^2,q^2)-
                             f_{\xi M_n^2}^2 (p^2,q^2) \right\} 
    \Bigg], \label{eq:kerK}
\end{eqnarray}
\begin{eqnarray}
f_M (p^2,q^2) 
&=& \frac{2}{p^2+q^2+M+\sqrt{(p^2+q^2+M)^2 -4p^2q^2}}, 
\end{eqnarray}
\begin{eqnarray}
\alpha_n &\equiv& \alpha \chi_n^2(y^\ast)
\qquad (n \ne 0), \\ 
\alpha_0 &=& \alpha = g^2/4\pi. 
\end{eqnarray}
Since our analysis is based on the effective theory on the 
four-dimensional brane below the {\it reduced} cutoff,
the explicit UV cutoff $\Lambda$ in terms of the 
loop momentum {\it on the brane} appears. 
$A(x)$ and $B(x)$ are the wave function normalization 
factor and the mass function of the fermion respectively. 
The mass function $B(x)$ is the oder parameter 
of chiral symmetry breaking on the brane. 
The chiral symmetry is broken down for $B(x) \ne 0$. 
By solving Eqs.~(\ref{eq:SDA}) and (\ref{eq:SDB}) 
we obtain the behavior of the dynamical fermion 
mass on the brane.

\section{Improved Ladder Analysis of Dynamical Mass on the Brane}
\label{sec:ila}

In this section we numerically solve the SD equation 
and analyze dynamical brane fermion mass induced by 
the bulk gauge theory. 
In the case of the bulk Abelian gauge theory with 
$N_{\rm KK} \lesssim 10$ we can use the point vertex 
approximation and the results are shown in \cite{Abe:2001yi}. 
In this section we extend the analysis in \cite{Abe:2001yi} 
to the bulk Yang-Mills theory. 
To solve the SD equation of Yang-Mills theory we use the 
improved ladder approximation where the running coupling 
is imposed to the vertex function.
The running coupling is 
derived from the truncated KK approach proposed in 
Ref.~\cite{Dienes:1998vh}. 
We use the asymptotic form (\ref{eq:KKmassasym}) of the KK 
spectrum to evaluate the running coupling in RS brane world. 

First we review the process of analyzing SD equation in 
usual 4D QCD \cite{Aoki:1990eq}, 
and then we extend it to the brane world bulk 
Yang-Mills theory. For convenience we choose bulk QCD 
($SU(3)_c$ Yang-Mills theory with $N_{\rm f}=3$ flavor fermion 
on the brane) as 
a concrete example because its results can be compared with 
the usual 4D QCD. We expect that the essential point of
the results is not changed in the other bulk Yang-Mills 
theory such as techni-color. 

\subsection{QCD on the Brane (Usual 4D QCD)}
Taking $N_{\rm KK}=0$ in Eqs.~(\ref{eq:SDA}) and (\ref{eq:SDB}),
the SD equation for usual QCD in four dimensional space-time 
(i.e. QCD on the brane) is obtained.
In this case the integration kernels (\ref{eq:kerL}) 
and (\ref{eq:kerK}) become 
\begin{eqnarray}
L_\xi (p^2,q^2;M_0=0,\alpha_0=\alpha)
 &=& \frac{\alpha}{4\pi} \,\xi\, q^2\, f_0^2 (p^2,q^2), 
     \label{eq:QCDkernelL} \\
K_\xi (p^2,q^2;M_0=0,\alpha_0=\alpha)
 &=& \frac{\alpha}{4\pi} \,(3 +\xi)\, f_0(p^2,q^2), 
     \label{eq:QCDkernelK}
\end{eqnarray}
where 
\begin{eqnarray}
f_0 (p^2,q^2) 
 &=& \theta (p^2-q^2) \frac{1}{p^2} + 
     \theta (q^2-p^2) \frac{1}{q^2}.
\end{eqnarray}
We usually use Landau gauge $\xi=0$ that results in 
$L_\xi (p^2,q^2;M_0=0,\alpha_0=\alpha)=0$, i.e. 
$A(p^2)=1$ in the SD equation. In that case the SD equation 
becomes single equation only with $B(p^2)$. 
$A(p^2)=1$ satisfies the QED like Ward-Takahashi identity.

We put the running coupling on the vertex function in the SD equation 
to compensate for neglecting three and four point gluon self 
interactions ${\cal L}_{\rm SI}$ in the ladder approximation. 
That is so called improved ladder approximation. 
We use the 1-loop perturbative running coupling,
\begin{eqnarray}
\alpha \equiv \frac{\pi}{3C_2(F)} \lambda (k^2), 
\end{eqnarray}
where
\begin{eqnarray}
\lambda (z) &=& \frac{\lambda_0}{1+\lambda_0B \ln \frac{z}{\mu^2}}, 
\end{eqnarray}
and $B = \frac{4\pi^2}{3C_2(F)}\beta_0$. We define $\beta_0$ by 
$\beta (g) = -\beta_0 g^3 + {\cal O}(g^5)$ and 
$\beta_0 = \frac{1}{16\pi^2} \big[ \frac{11}{3}C_2(G)-\frac{4}{3}N_{\rm f} T(F) \big]$ 
where $G$ and $F$ stand for the gauge group and the fermion representation 
under the group respectively. $C_2(G)$, $C_2(F)$ and $T(F)$ are 
defined by 
$\sum_{c,d}f_{acd}f_{bcd} \equiv C_2(G) \delta_{ab}$, 
$\textrm{tr} (T^a T^b) \equiv T(F) \delta_{ab}$ and 
$\sum_a T^a T^a \equiv C_2(F)1$. 
In the case that $G=SU(N=3)$ and $F$ is fundamental representation, 
these values are given by
\begin{eqnarray}
&&C_2(G)=N=3,\quad T(F)=\frac{1}{2}, \quad C_2(F)=\frac{N^2-1}{2N}=4/3, \\ 
&&B=\frac{1}{12C_2(F)}\left( \frac{11N-2N_{\rm f}}{3} \right)
 =\frac{9}{16}, \quad (N=3,N_{\rm f}=3). 
\end{eqnarray}

We also define $\Lambda_{\rm QCD}$, the dynamical scale of QCD, as 
\begin{eqnarray}
\frac{1}{\lambda (z)} - B \ln z = 
  \frac{1}{\lambda_0} - B \ln \mu^2 
  \ \equiv \ -B \ln \Lambda_{\rm QCD}^2, 
     \qquad \left( \Lambda_{\rm QCD}^2 \equiv 
     \mu^2/e^{\frac{1}{\lambda_0 B}} \right),
\end{eqnarray}
and obtain 
\begin{eqnarray}
\lambda (z) &=& \frac{16}{9}\frac{1}{\ln (z/\Lambda_{\rm QCD}^2)}. 
\label{eq:runlambda}
\end{eqnarray}
Following the analysis in Ref.~\cite{Miransky:vj} we introduce 
IR cutoff $z_{\rm IF}$ in the running coupling as 
\begin{eqnarray}
\alpha (z) &=& 
  \frac{\pi}{3C_2(F)} \left[
  \theta (z_{\rm IF}-z) \lambda (z_{\rm IF}) +
  \theta (z-z_{\rm IF}) \lambda (z)
  \right], 
\end{eqnarray}
and we take the following approximation
\begin{eqnarray}
\alpha &\to& \alpha (z) \simeq \alpha (x+y),
\label{eq:HMapprox} 
\end{eqnarray}
in the integration kernel (\ref{eq:QCDkernelL}) and 
(\ref{eq:QCDkernelK}). It is called Higashijima-Miransky 
approximation. This is due to the consideration that 
the mean value of the $\theta$-dependent part in $\alpha (z)$ 
where $z=x+y-\sqrt{xy}\cos \theta$ has only a negligible effect
after the angle integration in the SD 
equation\footnote{In the usual QCD analysis we introduce 
further approximation $\alpha (x+y) \simeq \alpha (\max [x,y])$, 
but we don't use it here.}. 
It is known that the approximation (\ref{eq:HMapprox}) 
violates the chiral Ward-Takahashi identity\footnote{As suggested 
in the second paper in Ref.~[9], we can keep the chiral Ward-Takahashi 
identity to use a non-local gauge.}. 
In the present paper we choose the gauge parameter $\xi$ to nearly 
hold the chiral Ward-Takahashi identity. 
In the following numerical analyses we use the experimental value 
\begin{eqnarray}
\Lambda_{\rm QCD} \simeq 200 \textrm{ MeV}, 
\end{eqnarray}
and choose $\ln (z_{\rm IF}/\Lambda_{\rm QCD})=1$ 
for the IR cutoff\footnote{To obtain the realistic pion decay 
constant in QCD in the ladder approximation, we need more 
careful treatment for the value of 
$\Lambda_{\rm QCD}$ and $z_{\rm IF}$ \cite{Aoki:1990eq}. 
In this paper we don't touch the detailed value of the results 
but the order of them.}. 

The composite Nambu-Goldstone field corresponds to the 
composite Higgs field in the dynamical electroweak symmetry 
breaking scenario. Its decay constant is obtained by the 
Pagels-Stokar approximation \cite{Pagels:1979hd},
\begin{eqnarray}
f_\pi^2 
= \frac{N}{4\pi^2} \int_0^{\Lambda^2} xdx
  \frac{B(x)\left( B(x)-xB'(x)/2 \right)}{\left(A^2(x)x+B^2(x)\right)^2}. 
\label{eq:PSformula}
\end{eqnarray}

\begin{figure}[htbp]
\begin{center}
\begin{minipage}{0.48\linewidth}
   \centerline{\epsfig{figure=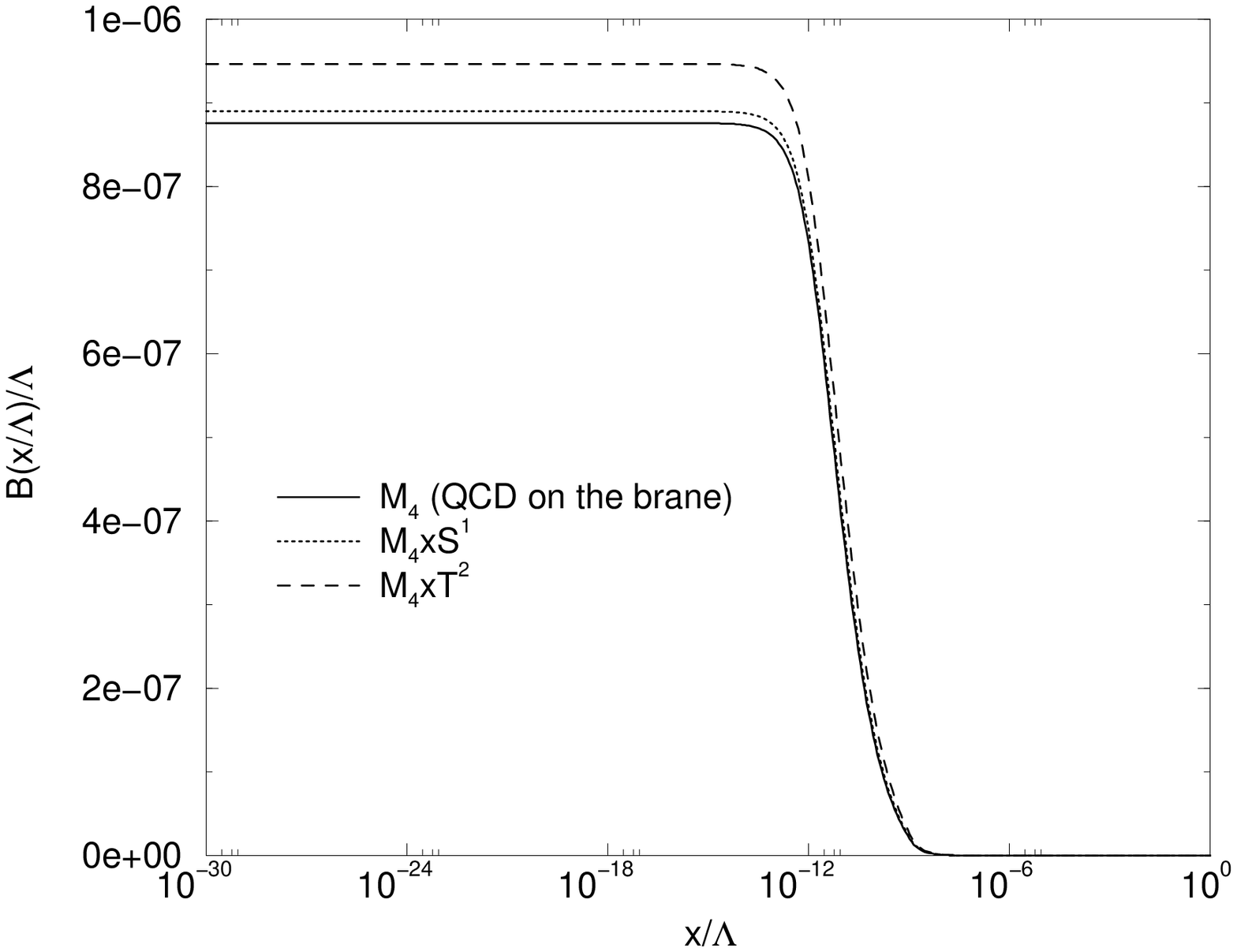,width=\linewidth}}
\end{minipage}
\hfill
\begin{minipage}{0.48\linewidth}
   \centerline{\epsfig{figure=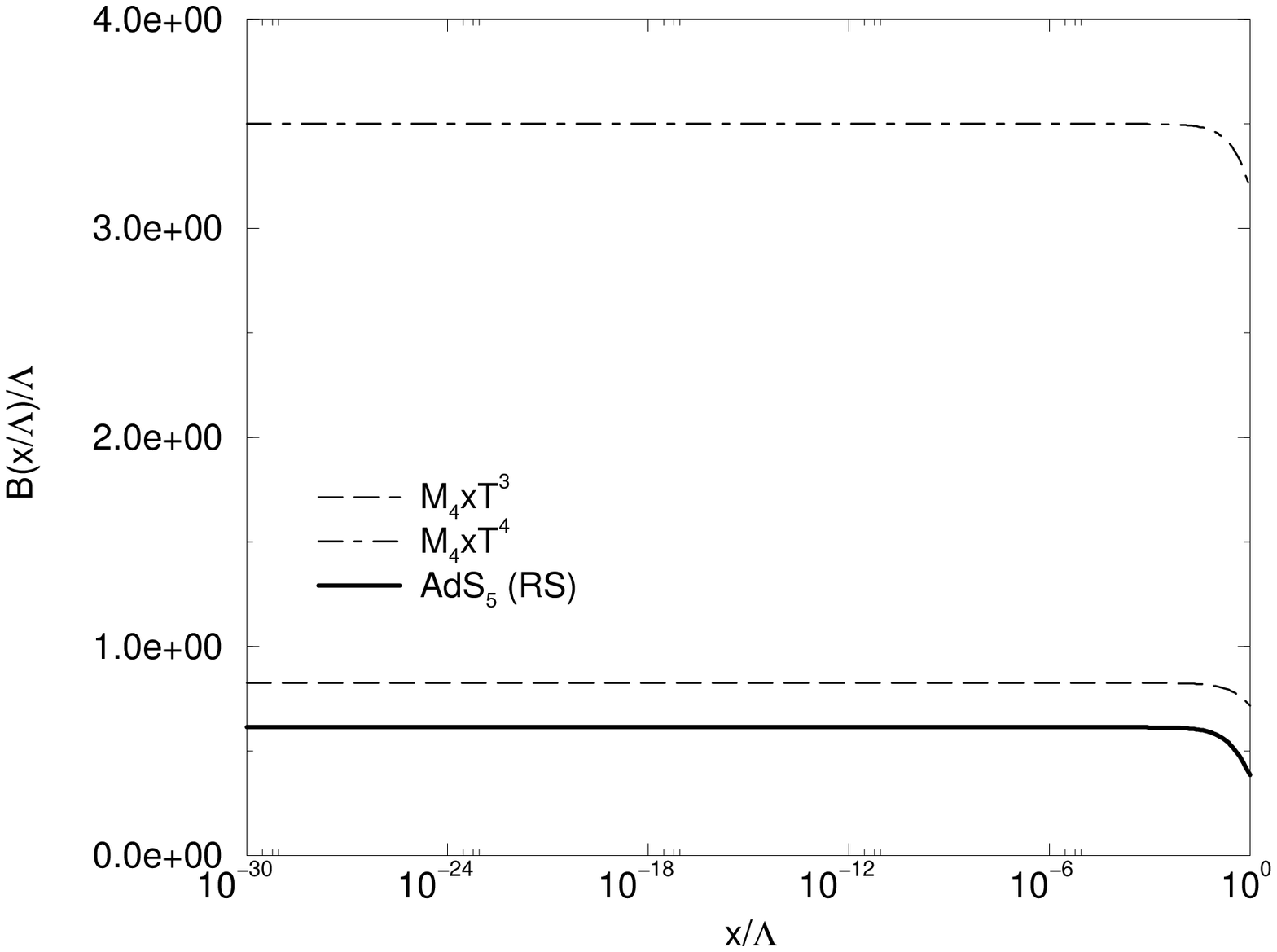,width=\linewidth}}
\end{minipage}
\end{center}
\caption{The behavior of $B(x)$ with various bulk space-time 
derived from the improved ladder SD equation. 
The case of (4D usual) QCD, bulk QCD in the flat brane world 
with $(4+\delta)$-dimensional bulk space-time ($\delta=1,2,3,4$) 
and in the RS brane world are shown. 
We set $\Lambda_{\rm QCD}=200$ MeV for all cases. 
The cutoff is taken as $\Lambda=10$ TeV for QCD and flat case, 
and for the flat brane world we take universal radii 
$R\Lambda_{\rm (4+\delta)D}=1$. 
For the RS brane world we take $k=\Lambda_{5D}=M_{\rm Pl}=10^{16}$ TeV
and $\Lambda=\pi e^{-\pi kR}\Lambda_{5D}=10$ TeV $(kR=11.35)$.}
\label{fig:SDB}
\begin{center}
\begin{minipage}{0.48\linewidth}
   \centerline{\epsfig{figure=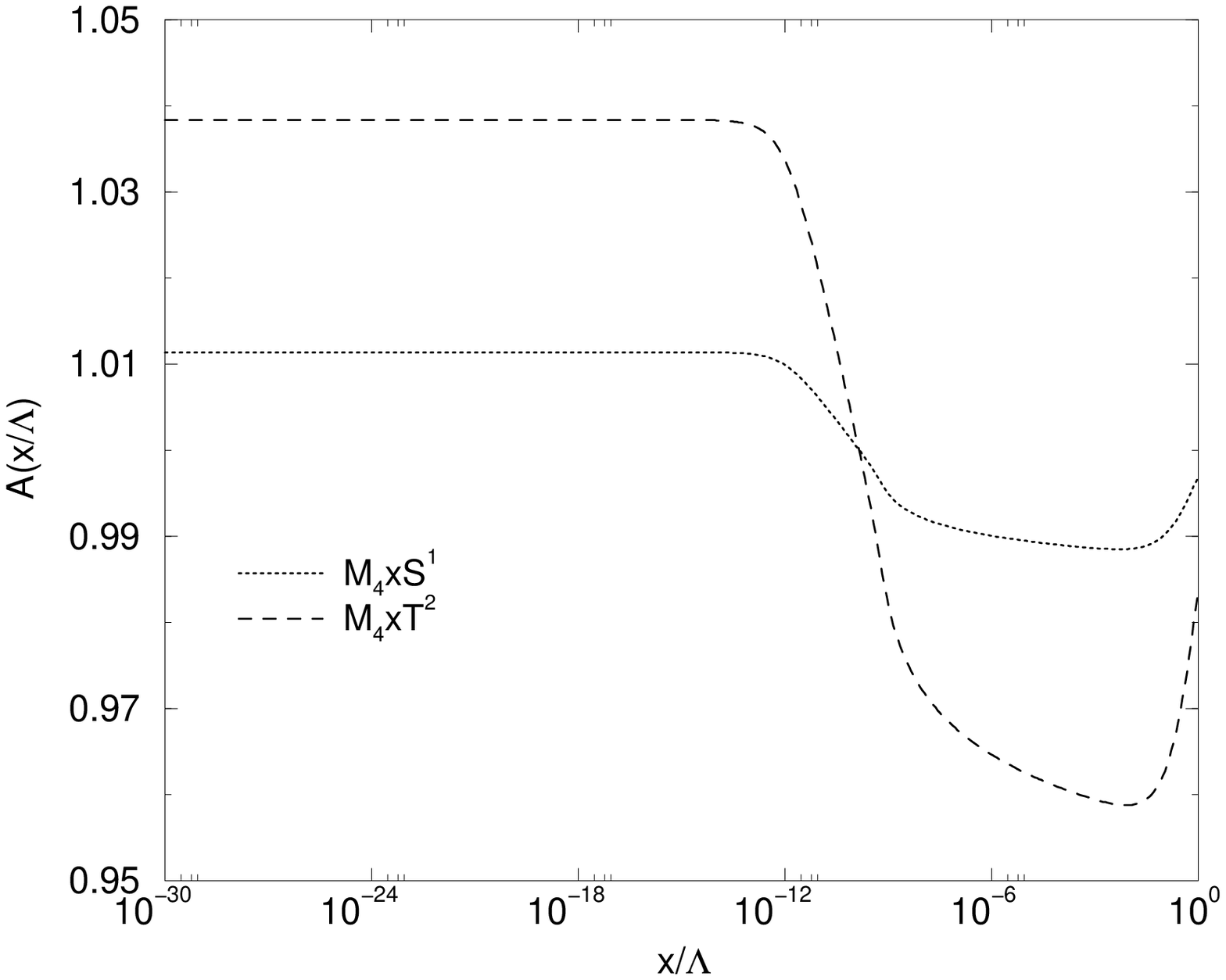,width=\linewidth}}
\end{minipage}
\hfill
\begin{minipage}{0.48\linewidth}
   \centerline{\epsfig{figure=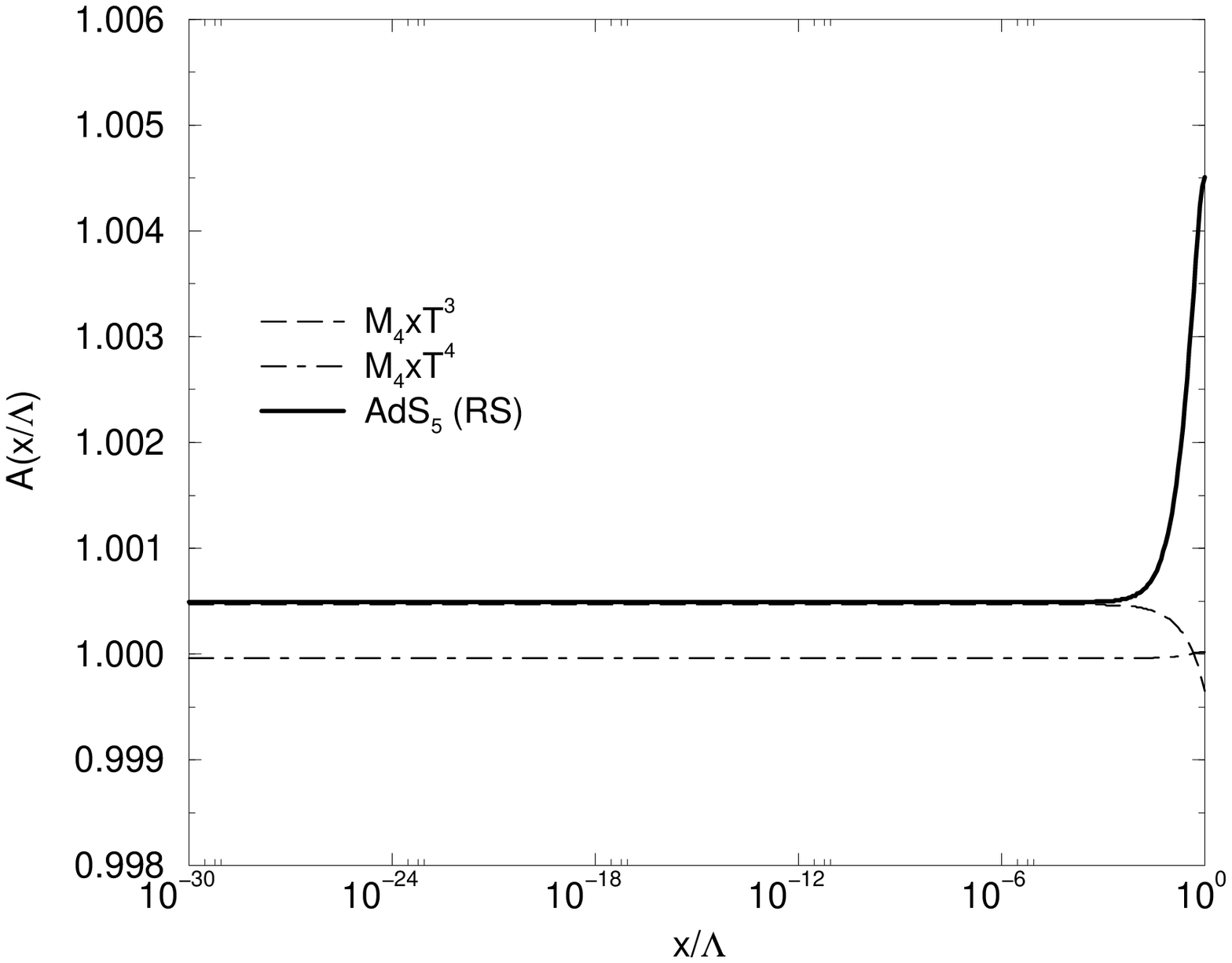,width=\linewidth}}
\end{minipage}
\end{center}
\caption{The behavior of $A(x)$ with various bulk space-time 
derived from the improved ladder SD equation. 
The case of (4D usual) QCD, bulk QCD in the flat brane world 
with $4+\delta$ dimensional bulk space-time ($\delta=1,2,3,4$) 
and in the RS brane world are shown. 
The mass parameters are taken as the same in the caption in 
Fig.~\ref{fig:SDB}. 
In order to satisfy Ward-Takahashi identity, i.e. $A(x) \simeq 1$ 
within $\pm 5$\%, we take tha gauge fixing parameter $\xi=0$ for QCD, 
$\xi=0.02$, $0.07$, $0.57$ and $0.63$ for flat brane world with 
$\delta=1$, $2$, $3$ and $4$ respectively and $\xi=0.50$ for RS case.}
\label{fig:SDA}
\end{figure}

\begin{figure}[t]
\begin{center}
\begin{minipage}{0.48\linewidth}
   \centerline{\epsfig{figure=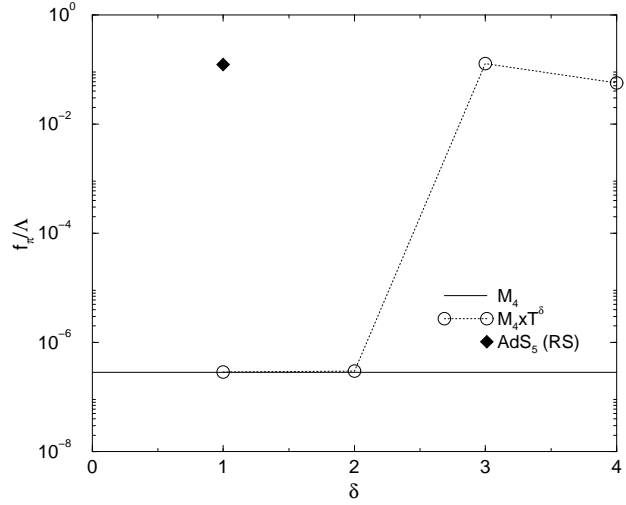,width=\linewidth}}
\end{minipage}
\end{center}
\caption{The behavior of the decay constant $f_\pi$ of the composite 
scalar field with various bulk space-time derived from the Pagels-Stokar formula. 
The case of (4D usual) QCD, bulk QCD in the flat brane world 
with $(4+\delta)$-dimensional bulk space-time ($\delta=1,2,3,4$) 
and in the RS brane world is shown together. 
The parameters are taken as the same in the caption in Fig.~\ref{fig:SDB}.}
\label{fig:fpi}
\end{figure}

Here we show the numerical results of the improved ladder SD 
equation in 4D QCD. We set the cutoff scale $\Lambda=10$ TeV 
in order to compare with the result of the brane world models. 
By using the replacement (\ref{eq:HMapprox}) 
we solve the SD equation (\ref{eq:SDA}), (\ref{eq:SDB}) 
and calculate the mass function 
$B(x)$. The behavior of the mass function $B(x)$ is shown 
as the thin solid line in Fig.~\ref{fig:SDB}. 
The scale of the fermion mass function is a little bit smaller
than the QCD scale, 
$B(x=\Lambda_{\rm QCD}^2) \sim {\cal O}(\Lambda_{\rm QCD}/10)$. 
Substituting the obtained mass function $B(x)$ to 
Eq.~(\ref{eq:PSformula}) we obtain the decay constant of the 
composite scalar field. As is shown with the thin solid line 
in Fig.~\ref{fig:fpi}, the decay constant of 4D QCD is also 
smaller than the QCD scale, $f_\pi \sim {\cal O}(\Lambda_{\rm QCD}/100)$. 
Since it is the result of the ordinary 4D QCD, the decay constant
is too small as Higgs field to break electroweak symmetry. 


\subsection{Bulk QCD in Flat Brane World}
Now we analyze the dynamical fermion mass induced 
by the bulk Yang-Mills theory in flat brane world, 
$M_4 \times T^\delta$. For simplicity we set
the same radii for all extra directions, i, e,
 universal extra dimensions. We study $SU(N=3)$ bulk Yang-Mills 
theory (QCD) with $N_f=3$ flavor fermion as a concrete example.
For QCD extended in the compact extra dimension with the radius $R$, 
the running coupling $\alpha (z)$ has the power low 
behavior \cite{Dienes:1998vh} in the region $1/R^2 \le z$, 
\begin{eqnarray}
\alpha^{-1} (z)
&=& \alpha^{-1}_0 
   +2\pi \beta_0 \ln \frac{z}{\mu_0^2} 
   -2\pi \tilde{\beta}_0 \ln \frac{z}{\mu_R^2}
   +2\pi \tilde{\beta}_0 \frac{2X_\delta}{\delta}
    \left[ \left(\frac{z}{\mu_R^2}\right)^{\delta/2} -1 \right], 
\end{eqnarray}
where $\mu_R=1/R$ and 
\begin{eqnarray}
X_\delta 
&=& \frac{\pi^{\delta/2}}{\Gamma (1+\delta/2)} 
 =  \frac{2\pi^{\delta/2}}{\delta \Gamma (\delta/2)}, \quad 
\alpha_0 \equiv \alpha (\mu_0^2), 
\end{eqnarray}
and $\tilde{\beta}_0$ is the beta-function coefficient 
stimulating from the bulk fields. Thus we obtain the running 
coupling of the bulk QCD with $N_{\rm f}=3$ flavor fermion on the brane,
\begin{eqnarray}
\lambda (z)
&=& \frac{1}
    {B \ln (z/\Lambda_{\rm QCD}^2) 
      -\widetilde{B} 
       \left[ \ln (z/\mu_R^2) 
      -\frac{2X_\delta}{\delta} 
        \left\{ (z/\mu_R^2)^{\delta/2}-1 \right\}
        \right] \theta (z-\mu_R^2)}, 
\label{eq:powerlow}
\end{eqnarray}
where 
\begin{eqnarray}
\left\{
\begin{array}{l}
B=\frac{1}{24C_2(F)}\left( \frac{11N-2N_{\rm f}}{3} \right)
 =9/16 \quad (N=3,N_{\rm f}=3) \\
\widetilde{B} 
 =\frac{1}{24C_2(F)}\left( \frac{11N-2\tilde{N}_{\rm f}}{3} \right)
 =11/16 \quad (N=3,\tilde{N}_{\rm f}=0) 
\end{array}
\right.. \label{eq:BandtB}
\end{eqnarray}

Using the running coupling (\ref{eq:powerlow}), we numerically
solve the improved ladder SD equation and find the behaviors 
of the fermion wave function $A(x)$ and the mass function  $B(x)$ 
on the brane. We analyze the SD equation on the 
$(y_1^\ast,\ldots,y_\delta^\ast)$ $=(0,\ldots,0)$ brane, 
with $\Lambda=\Lambda_{\rm (4+\delta)D}=10$ TeV, $\Lambda_{\rm QCD}=200$ MeV 
and $R\Lambda_{\rm (4+\delta)D}=1$ 
as typical cases, where $\Lambda_{\rm (4+\delta)D}$ and $\Lambda$ are 
the bulk fundamental scale and the reduced cutoff scale on the brane 
respectively. For $\delta=1,2,3$ and $4$, the behavior of 
$B(x)$ and $A(x)$ are drawn in Figs.~\ref{fig:SDB} and \ref{fig:SDA} 
respectively. For $\delta<3$ the fermion mass function behaves
like 4D QCD, i.e. 
$B(x=\Lambda_{\rm QCD}^2) \sim {\cal O}(\Lambda_{\rm QCD}/10)$.
For $\delta \ge 3$ the situation is dramatically changed. 
The fermion mass function develops a value near 
the {\it reduced} cutoff scale, 
$B(x=\Lambda_{\rm QCD}^2) \sim {\cal O}(\Lambda)$.
Using the Pagels-Stokar formula (\ref{eq:PSformula}),
we calculate the decay constant of the composite scalar field. 
As is shown in Fig.~\ref{fig:fpi} the scale of the decay constant 
is summarized as $f_\pi \sim {\cal O}(\Lambda_{\rm QCD}/10)$ 
for $\delta<3$ and $f_\pi \sim {\cal O}(\Lambda/10)$ for $\delta \ge 3$. 
Therefore if the bulk space-time has seven or more dimensions, 
the bulk Yang-Mills theory induces the TeV scale decay constant on 
the brane. There is a possibility that the bulk QCD or techni-color 
can provide a composite Higgs field in the flat brane world. 
In this section we take 
$1/R = \Lambda = \Lambda_{\rm (4+\delta)D} = 10$ TeV. 
We comment that this setup merges into the scenario of 
`large extra dimension' \cite{Antoniadis:1990ew} in which the 
fundamental scale is around TeV and we assume the existence of 
a large extra dimension where only the gravity propagates.

\subsection{Bulk QCD in RS Brane World}

Here we study the bulk QCD in the RS brane world.
In the warped brane world a mass scale 
$\Lambda$ on the brane at $y=y^\ast$ is suppressed by 
the warp factor, i.e. 
$\Lambda \simeq W(y^\ast) \Lambda_{\small 5D}$, 
where $\Lambda_{\small 5D}$ is the fundamental scale of the theory 
and $W(y)=e^{-\pi k|y|}$ is the warp factor. 
As is discussed in the previous subsection, it is natural to 
consider that the effective theory of the brane fermion should be 
regularized by the {\it reduced} fundamental scale 
{\it on the brane}. 
Hence we cutoff the loop momentum in the SD equation at the {\it reduced} 
cutoff scale $\Lambda$ on the brane,
\begin{eqnarray}
\Lambda \simeq e^{-k|y^\ast|}\Lambda_{\small 5D} 
        \simeq e^{-k|y^\ast|}k
        \qquad (y^\ast =0,\pi R). 
\end{eqnarray}
For $kR \simeq 11$ UV cutoff scale $\Lambda$ is suppressed by 
$10^{-15}$ on the $y^\ast=\pi R$ brane. Randall and Sundrum
obtain the weak-Planck hierarchy form this warp suppression
factor in the original RS model \cite{Randall:1999ee}. 

The behavior of the running coupling is non-trivial 
in the RS space-time. 
In this paper we evaluate it in terms of the 4D effective 
theory below the cutoff scale $\Lambda$ by using the truncated 
KK approach. 
In the region $kR \gg 1$ and $M_n \ll k$, KK mass spectrum of
the gauge boson reduces to the asymptotic form (\ref{eq:KKmassasym}). 
Comparing it with the torus case (\ref{eq:KKmasstorus}), we find that 
the running coupling in the RS space-time is obtained 
by replacing $\mu_R$ with $\mu_{kR}$ and shifting the KK lowest 
threshold by $\mu_{kR}/4$. In addition we must take the orbifold 
condition into account. It drops all the $Z_2$ odd modes. 
Only a half contribution comes from the KK excited modes.
Thus we replace $\widetilde{B} \to \widetilde{B}/2$ in 
Eq.~(\ref{eq:powerlow}). The running coupling reads
\footnote{We obtain power low running from the truncated KK 
approach. In Refs.~\cite{Randall:2001gb} the running coupling of 
the bulk gauge theory  is investigated recently in the warped 
background. These papers insist that the running is only
logarithmic. If we use it, the effect of the KK modes 
will be enhanced because the power low suppression is not 
exist in the integrand in SD equation.} 
\begin{eqnarray}
\lambda (z)
  = \frac{1}
    {B \ln (z/\Lambda_{\rm QCD}^2) 
      -(\widetilde{B}/2) 
       \left[ \ln (\sqrt{z}/\mu_{kR} + 1/4)^2 
      -4\left\{ \sqrt{z}/\mu_{kR} - 3/4 \right\}
        \right] \theta (z-(3/4)^2\mu_{kR}^2)}, 
\label{eq:RSpowerlow}
\end{eqnarray}
where $B$ and $\widetilde{B}$ are given by Eq.~(\ref{eq:BandtB}). 

Applying the truncated running coupling (\ref{eq:RSpowerlow})
we evaluate the SD equation on the $y=\pi R$ brane, with 
$k=\Lambda_{5D}=M_{\rm Pl}=10^{16}$ TeV,
$\Lambda=\pi e^{-\pi kR}\Lambda_{5D}=10$ TeV and 
$\Lambda_{\rm QCD}=200$ MeV as typical cases.
The solution of $B(x)$ and $A(x)$ are plotted in Figs.~\ref{fig:SDB} 
and \ref{fig:SDA} respectively. 
Calculating the Eq.~(\ref{eq:PSformula}) we obtain the decay 
constant of the composite scalar as is shown in Fig.~\ref{fig:fpi}. 
In the RS brane world both scales of the fermion mass function 
and the decay constant are near the {\it reduced} cutoff scale,
$B(x=\Lambda_{\rm QCD}^2) \sim {\cal O}(\Lambda/10)$, 
$f_\pi \sim {\cal O}(\Lambda/10)$ $\sim$ ${\cal O}(1)$ TeV. 
Therefore the bulk Yang-Mills theory induces TeV scale decay 
constant on the brane. It means that the bulk QCD has possibility
to realize the composite Higgs scenario of electroweak symmetry
breaking. 

The original theory has only the Planck scale, it has no TeV scale 
source but the fermion mass and decay constant of the composite 
scalar are generated at TeV scale through the warp suppression of 
the {\it reduced} cutoff scale on the $y=\pi R$ brane. 
KK excited modes of the gauge field are localized at the $y=\pi R$ 
brane in RS model. It enhances the contribution from the KK modes
and realize the TeV scale composite scalar on the brane with
only one extra dimension. This is the difference from the 
(4+1)-dimensional flat brane world, $\delta=1$. 
The bulk gauge theory with only Planck scale in RS brane world can 
acquire a TeV scale dynamically due to a fermion pair condensation 
on the $y=\pi R$ brane through the effect of the gauge boson KK excited 
modes localized at the brane. 
The warp suppression mechanism in the RS model is realized 
dynamically.

The extension of above results to the {\it bulk SM gauge fields} 
will make us to propose two types of primary models of dynamical 
electroweak symmetry breaking. 
We call one of them {\it (1,2)-3 model} in which the 
first and second generations in the SM are confined on the $y=0$ 
brane, and the third generation is on the $y=\pi R$ brane. 
The KK excited mode is not localized on the $y=0$ brane shown in 
Fig.~\ref{fig:WF}. The coupling between the KK excited 
gluons and the brane fermion are suppressed by the mode function
$\chi_n (y=0)$ on the $y=0$ brane.
Therefore we expect that the top quark-pair is condensed on 
the $y=\pi R$ brane, while the up and charm quark are not 
condensed.
The top mode SM \cite{Miransky:1988xi} can be realized in 
RS brane world \cite{Rius:2001dd}. 
We note that the zero modes function, $\chi_0$, are constant, so 
that the fermions on the both brane equally 
couples to the massless gauge filed (see also Fig.~\ref{fig:WF}). 
We can construct the other model, {\it (1,2,3)-4 model}, which
is defined as follows. All the SM fermions lie in $y=0$ brane 
and the fourth generation (or the techni-fermion) lies 
in the $y=\pi R$ brane. In this model 
the fourth generation can develops TeV scale mass function 
and decay constant by KK excited gluons (or techni-gluons).

It is expected that these models can provide composite Higgs 
and the TeV scale physics simultaneously. Although some 
problems are remained, e.g. how to dynamically realize the 
detailed mass relation (dynamical Yukawa couplings) in SM, 
or flavor breaking etc.. 
\begin{figure}[htbp]
\begin{center}
\begin{minipage}{0.48\linewidth}
   \centerline{\epsfig{figure=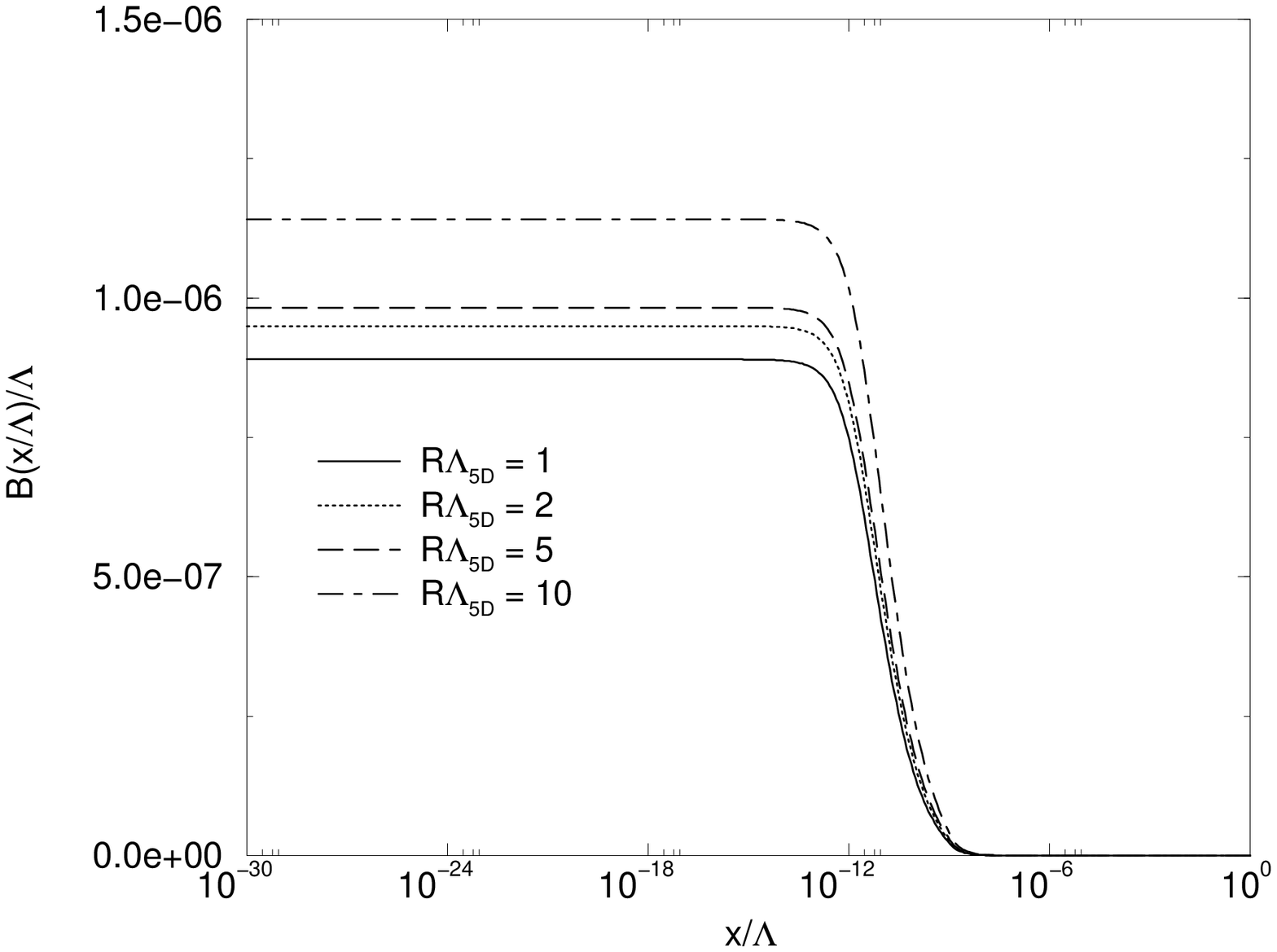,width=\linewidth}}
\end{minipage}
\hfill
\begin{minipage}{0.48\linewidth}
   \centerline{\epsfig{figure=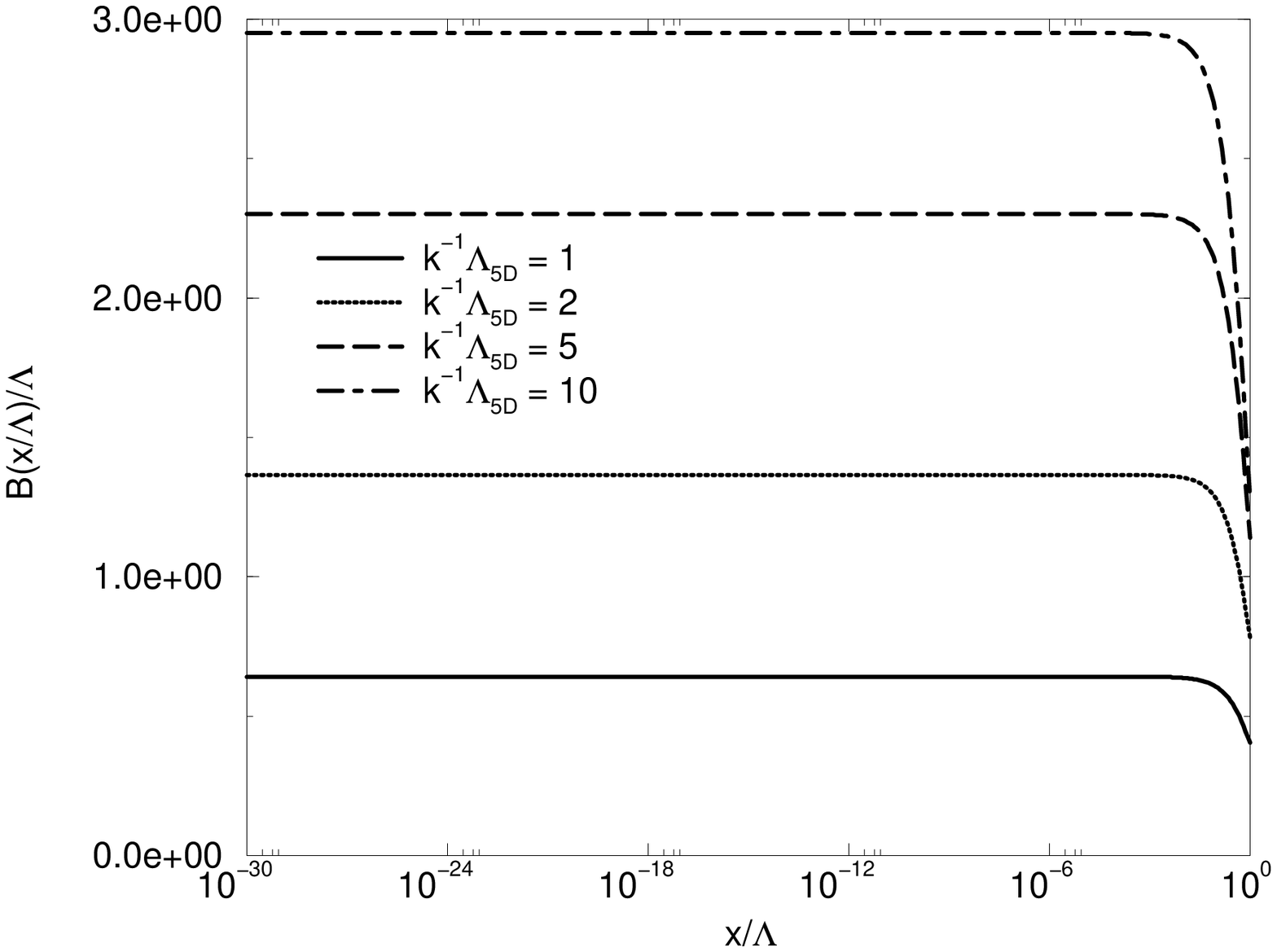,width=\linewidth}}
\end{minipage}
\end{center}
\caption{The behavior of $B(x)$ with $(4+1)$-dimensional bulk space-time 
derived from the improved ladder SD equation. 
The case of $(4+1)$-dimensional bulk QCD in the flat brane world 
with the radius $R\Lambda_{\rm 5D}$$=1,2,5$ and $10$, and the case 
in the RS brane world with the AdS$_5$ curvature radius 
$k^{-1}\Lambda_{\rm 5D}$$=1,2,5$ and $10$ are shown. 
We set $\Lambda_{\rm QCD}=200$ MeV for all cases. 
The cutoff is taken as $\Lambda=10$ TeV for flat case. 
For the RS brane world we take $k=\Lambda_{5D}=M_{\rm Pl}=10^{16}$ TeV
and $\Lambda=\pi e^{-\pi kR}\Lambda_{5D}=10$ TeV $(kR=11.35)$.}
\label{fig:SDBR}
\begin{center}
\begin{minipage}{0.48\linewidth}
   \centerline{\epsfig{figure=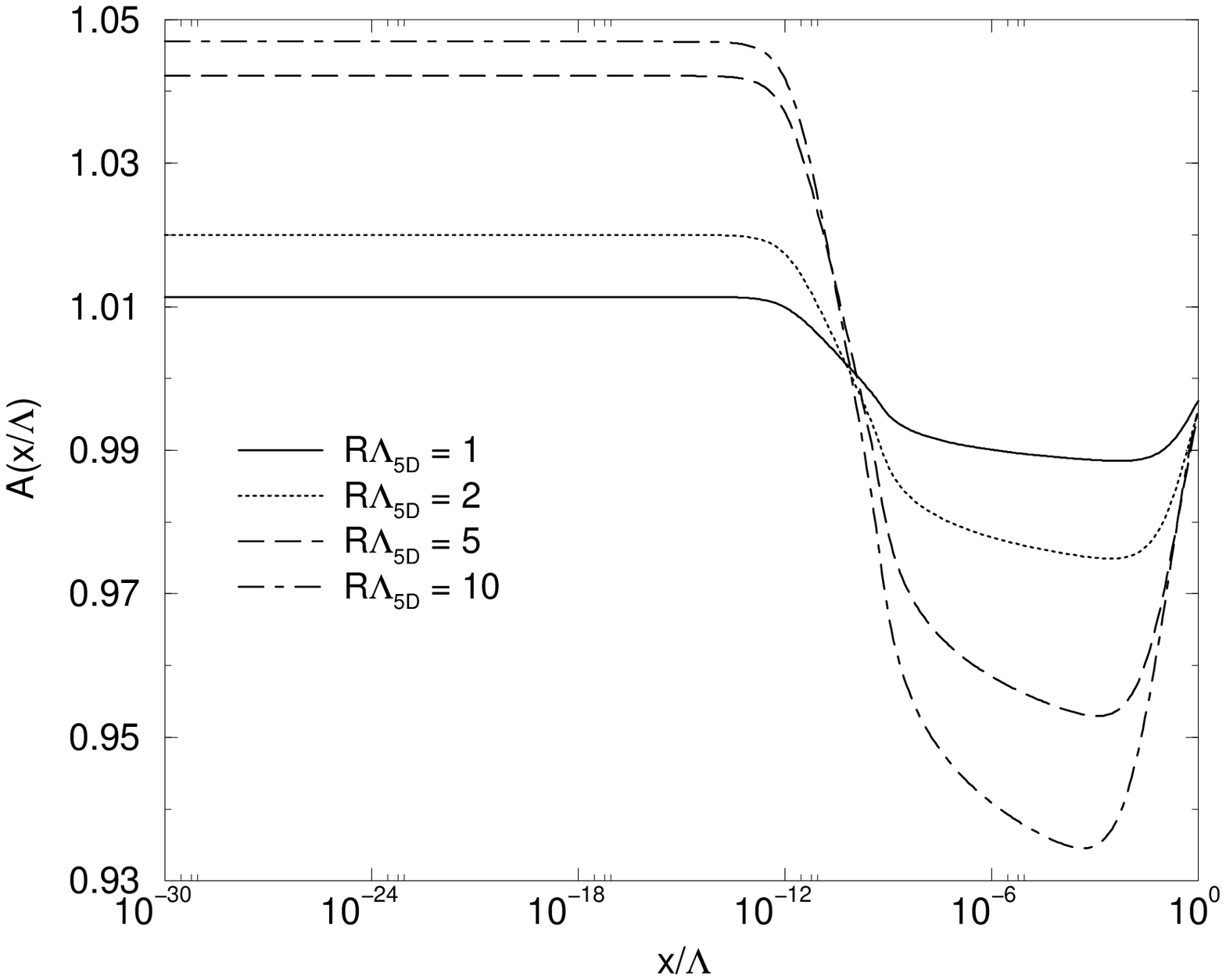,width=\linewidth}}
\end{minipage}
\hfill
\begin{minipage}{0.48\linewidth}
   \centerline{\epsfig{figure=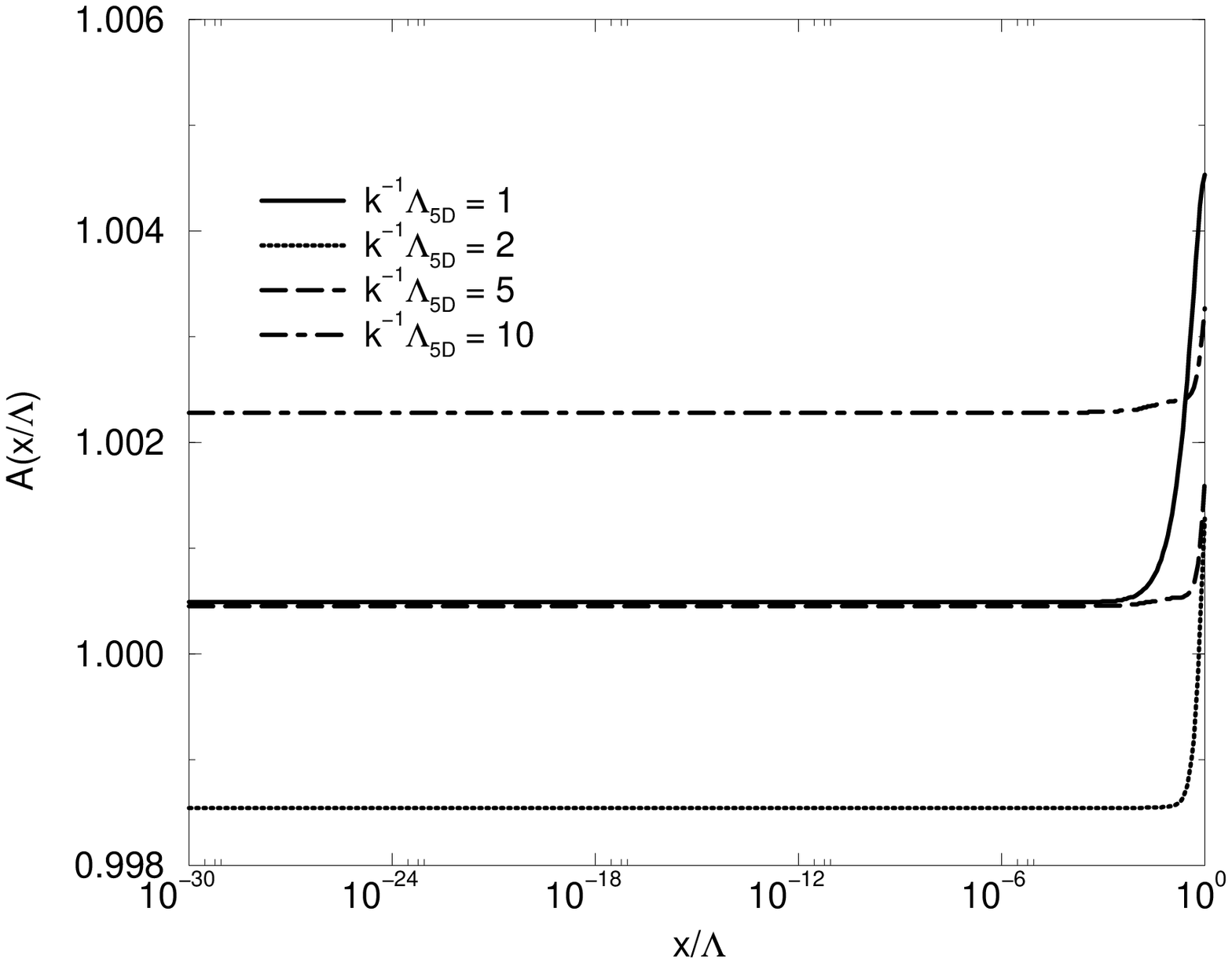,width=\linewidth}}
\end{minipage}
\end{center}
\caption{The behavior of $A(x)$ with $(4+1)$-dimensional bulk space-time 
derived from the improved ladder SD equation. 
The case of $(4+1)$-dimensional bulk QCD in the flat brane world 
with the radius $R\Lambda_{\rm 5D}$$=1,2,5$ and $10$, and the case 
in the RS brane world with the AdS$_5$ curvature radius 
$k^{-1}\Lambda_{\rm 5D}$$=1,2,5$ and $10$ are shown. 
The mass parameters are taken as the same in the caption in 
Fig.~\ref{fig:SDBR}. 
In order to satisfy Ward-Takahashi identity, i.e. $A(x) \simeq 1$ 
within $\pm 7$\%, we take tha gauge fixing parameter 
$\xi=0.020$, $0.040$, $0.080$ and $0.104$ for flat brane world with 
$R\Lambda_{\rm 5D}=1$, $2$, $5$ and $10$ respectively, and 
$\xi=0.35$, $0.37$, $0.41$ and $0.50$ for RS brane world with 
$k^{-1}\Lambda_{\rm 5D}=1$, $2$, $5$ and $10$ respectively.}
\label{fig:SDAR}
\end{figure}
\begin{figure}[t]
\begin{center}
\begin{minipage}{0.48\linewidth}
   \centerline{\epsfig{figure=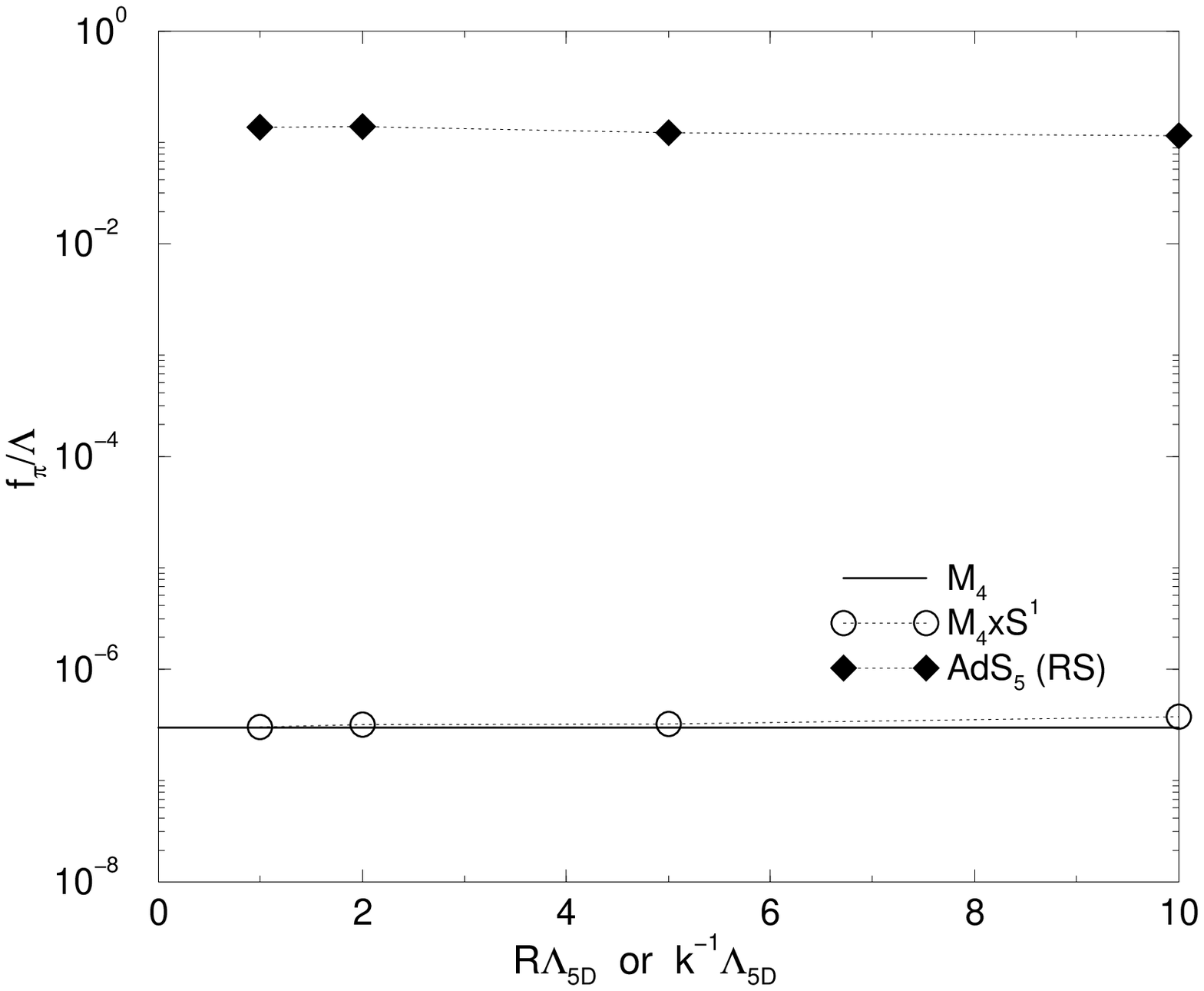,width=\linewidth}}
\end{minipage}
\end{center}
\caption{The behavior of the decay constant $f_\pi$ of the composite 
scalar field with $(4+1)$-dimensional bulk space-time 
derived from the Pagels-Stokar formula. 
The case of $(4+1)$-dimensional bulk QCD in the flat brane world 
with the radius $R\Lambda_{\rm 5D}$$=1,2,5$ and $10$, and the case 
in the RS brane world with the AdS$_5$ curvature radius 
$k^{-1}\Lambda_{\rm 5D}$$=1,2,5$ and $10$ are shown together. 
The parameters are taken as the same in the caption in Fig.~\ref{fig:SDBR}.}
\label{fig:fpiR}
\end{figure}

\subsection{Classification of $B(x)$ and $N_{\rm KK}$-Dependence}

The behaviors of $B(x)$ shown in Fig.~\ref{fig:SDB} are divided 
into two pieces. QCD in the flat bulk space-time with $\delta=1$ and $2$ is 
classified into {\it Yang-Mills type}, i.e. the contribution from
the KK modes is extremely small and $B(x)$ has a similar profile to
the ordinary QCD (QCD on the brane). On the other hand, 
flat bulk space-time with $\delta=3$, $4$ and the RS brane world 
are {\it NJL type}, i.e. KK modes contribution dominates the 
symmetry breaking. 

In Figs.~\ref{fig:SDB}, \ref{fig:SDA} and \ref{fig:fpi}, 
we consider only the case $N_{\rm KK}=1$. 
Finally we analyze the $R$- and $k$-dependence of the wave function
$A(x)$ and the mass function $B(x)$ on the brane embedded in 
(4+1)-dimensional bulk space-time. The behaviors of $A(x)$ and $B(x)$
are shown in Figs.~\ref{fig:SDBR}, \ref{fig:SDAR} and \ref{fig:fpiR}. 
There are little $R$- and $k$-dependence. Only a small 
$N_{\rm KK}$-dependence is observed in the behavior of $B(x)$.
Eqs.~(\ref{eq:KKmasstorus}) or (\ref{eq:KKmassasym}) show 
that the number of the KK mode $N_{\rm KK}$ increases as $R$ or 
$k^{-1}$ increases. The influence of the KK mode propagation
seems to be enhanced for larger $R$ or $k^{-1}$.
In Fig.~\ref{fig:SDBR} we clearly observe that the mass 
function $B(x)$ grows up as $R$ or $k^{-1}$ increases.
In Eqs.~(\ref{eq:powerlow}) or (\ref{eq:RSpowerlow}) the 
power low running terms contain the step function 
$\theta (z-\mu_R^2)$ or $\theta (z-(3/4)^2\mu_{kR}^2)$. 
Since the KK mode mass scale $\mu_R$ and $\mu_{kR}$ are 
proportional to $1/R$ and $k$ respectively, the contribution 
from these terms is enhanced as $R$ or $k^{-1}$ increases.
The contribution from the power low running suppresses 
dynamical symmetry breaking.
For example, if we drop the power low running term
in $(4+1)$-dimensional flat bulk space-time, $B(x)$ blows up 
from the {\it Yang-Mills type} to the {\it NJL type} as 
$N_{\rm KK}$ increases.
Thus we conclude that the bulk gluon self interactions 
${\cal L}_{\rm SI}$, i.e. the power low running coupling, 
act as a suppression factor for the dynamical mass on the 
brane.

\section{Conclusion}
\label{sec:concl}

We have studied the basic structure of dynamical symmetry breaking 
on the four dimensional brane in the bulk Yang-Mills theory (QCD). 
Using the 4D effective theory of the bulk QCD and the improved 
ladder SD equation, we have found that the dynamical mass scale 
can be affected by the {\it reduced} cutoff scales on the brane. 
Starting from the Yang-Mills theory in $M_4 \times T^\delta$ space-time 
which couples to a fermion on the four dimensional brane we derive 
four dimensional effective theory by KK reduction. 
We take $A_{3+i}=0$ $(i=1,\ldots,\delta)$ gauge in extra directions 
but leave the gauge fixing parameter free parallel to the brane. 
Our system needs a proper regularization because it is a theory beyond 
four dimensional space-time where the gauge theory is unrenomalizabele. 
We impose the cutoff regularization at the UV {\it reduced} 
scale $\Lambda$ on the brane fermion lives in. For the flat brane 
world the reduced scale should be equal to the fundamental scale 
in the original bulk Yang-Mills theory, i.e. 
$\Lambda \simeq \Lambda_{\rm (4+\delta)D}$. 
The RS brane world, however, has a reduced scale suppressed 
by the warp factor, that is $\Lambda \simeq e^{-k|y^\ast|}\Lambda_{5D}$ 
which depends on the position in the extra direction. 

Based on the 4D effective theory we derive the SD equation for the fermion 
propagator on the brane. From the four dimensional point of view the
equation corresponds to the simultaneous integral equation. The loop 
integral consists of one massless and $N_{\rm KK}$ massive vector field 
where $N_{\rm KK}$ is the number of the KK modes under the reduced cutoff 
scale. Using the iteration method we numerically solve it for typical 
values of radius, $R\Lambda_{\rm (4+\delta)D}=1$, and the (AdS) curvature 
scale, $k=\Lambda_{5D}$. In order to make the analysis consistent with 
the QED like Ward-Takahashi identity, we choose the appropriate 
value of the gauge fixing parameter for each analysis. 

The results of the numerical analysis is as follows. For a flat extra 
dimension $\delta=1$ the dynamical mass of the brane fermion 
is the same scale as the usual 4D QCD result, i.e. there is little
contribution from the KK excited gauge boson. We also study the case 
of the flat brane world with more than one extra dimension. The numerical 
analysis of the SD equation shows that the number of the extra dimension
must be no less than three, $\delta \ge 3$, to generate the TeV scale 
dynamical fermion mass on the brane. Therefore it is obtained in the case 
of the Yang-Mills theory in seven or more dimensional bulk space-time. 

The RS brane world has one extra dimension which corresponds to 
$\delta=1$ case but it is curved. Because the gauge boson KK excited 
modes localized at the $y=\pi R$ brane due to the curvature effect 
(see Fig.\ref{fig:WF}), we have a significant result. 
The dynamical mass of the brane fermion at the $y=\pi R$ is the order 
of the {\it reduced} cutoff scale that is warp suppressed from the 
fundamental scale.
We also estimate the decay constant of the composite Nambu-Goldstone scalar
field by using the Pagels-Stokar approximation. 
The decay constant is also warp suppressed on the $y=\pi R$ brane
in the RS brane world.
If the chiral symmetry is gauged as in the SM case, the composite
scalar absorbed in the longitudinal component of the gauge boson.
This decay constant is directly related to the mass of the gauge boson. 
The result in the RS brane world provides a possibility to realize 
the situation that electroweak symmetry breaking is triggered by 
the condensation of the composite Higgs field on the $y=\pi R$ brane.
Such a mechanism is caused by the bulk Yang-Mills theory, for example 
the techni-color or the QCD itself in the SM. Therefore we can solve 
the weak-Planck hierarchy from the warp factor, that is realized, in a 
sense, dynamically. This is one of the dynamical realization of so called 
Randall-Sundrum model. 

Our analyses are based on the 4D effective theory of the bulk gauge theory 
with an explicit UV cutoff in the loop momentum integration. 
We can also check the regularization dependence 
of our model by comparing the results of the lattice regularized 
(deconstructed) version. The deconstructed bulk field theory 
is given in \cite{Abe:2002rj} on the RS background. 

The results of our analysis are also important in a theoretical sense. 
We consider that the SD equation includes certain non-perturbative effect 
of the theory. The non-trivial results obtained here implies that 
the bulk gauge theory may have rich non-perturbative dynamics 
that depends on the configurations of the extra space.
In the brane world we should take account of such effect 
in addition to the perturbative analysis. 
Our results imply that the bulk gauge dynamics can induce various types 
of the {\it dynamical} symmetry breaking in the brane world, 
depending on the bulk-brane configuration. 
For instance, we may obtain the dynamical breaking of the grand unification 
gauge symmetry \cite{Kugo:1994qr} (or even supersymmetry?), 
in addition to the possibility of dynamical electroweak symmetry breaking 
emphasized in this paper. 
We are also interested in the consistency of these models with the realistic 
D-brane system in the string theory. 

\section*{Acknowledgements}
The authors would like to thank Kenji~Fukazawa,  Michio~Hashimoto, 
Tatsuo~Kobayashi and Masaharu~Tanabashi for fruitful discussions and 
correspondences. We also thank Kensaku~Ohkura for stimulating discussions. 

\pagebreak[0]

\end{document}